\documentclass[twocolumn,superscriptaddress,floatfix,preprintnumbers,aps,prl]{revtex4-1}
\usepackage{graphics,amssymb,amsmath,epsfig,color}
\usepackage{graphicx}
\usepackage{overpic}

\begin{document}

\title{Theory of the Josephson Junction Laser}
\author{Steven H. Simon}
\affiliation{Rudolf Peierls Centre, Oxford University, OX1 3NP, United Kingdom}
\author{Nigel R. Cooper}
\affiliation{T.C.M. Group, Cavendish Laboratory, J.J. Thomson Avenue, Cambridge, CB3 0HE, United Kingdom}
\date{\today}
\begin{abstract}
We develop an analytic theory for the recently demonstrated Josephson Junction laser (Science {\bf 355}, 939, 2017).  By working in the time-domain representation (rather than the frequency-domain) a single non-linear equation is obtained for the dynamics of the device, which is fully solvable in some regimes of operation.    The nonlinear drive is seen to lead to mode-locked output, with a period set by the round-trip time of the resonant cavity.
\end{abstract}
\maketitle

The physics of Josephson junctions has been studied intensely for over half a century\cite{Barone,Likharev}.  These devices are fairly unique as the only lossless nonlinear low temperature circuit elements\cite{Devoret1}.   Due to interest in using Josephson physics for quantum computing applications,  enormous attention has focused on the combination of Josephson junctions and low loss resonant cavities\cite{Devoret2,LesHouches}.     

\begin{figure}
	\vspace*{-.2in}
	\includegraphics[width=2.25in,angle=90]{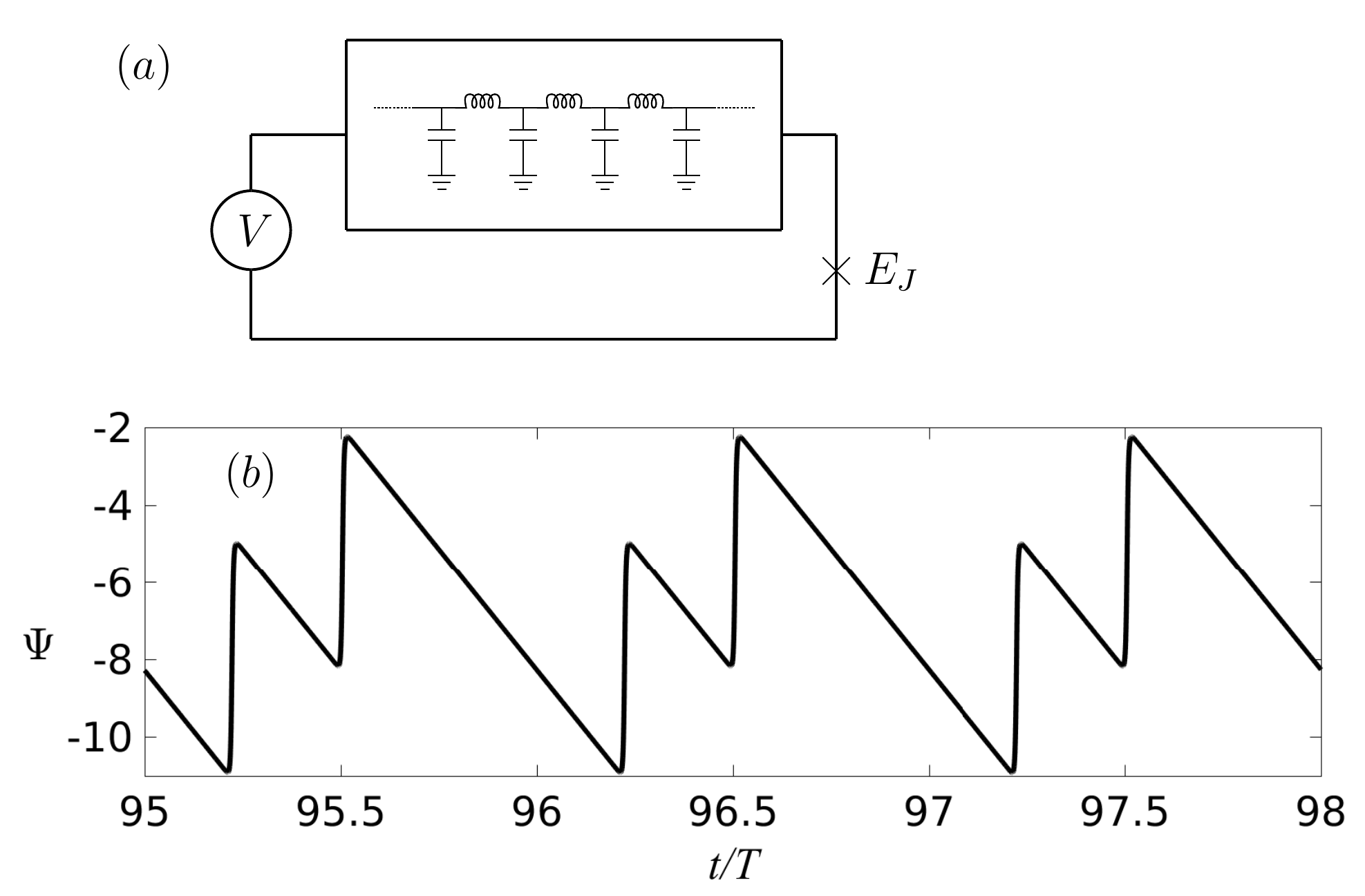}
	\vspace*{-.2in}
	\caption{(a) Circuit diagram for the Josephson junction laser (From Refs \cite{Cassidy,Armour,Meister}).    The cavity can be thought of as an LC chain (See Supplemental Material\cite{supplement}).  (b) Numerical form of $\Psi(t)$ obtained by integration of Eq.~\ref{eq:dynamics2} for parameters $V = 2, \gamma=.01, \lambda=6$ in units where $T=2 \pi$ so $\omega_0=1$.   The waveform has constant negative slope with discrete steps of $2 \pi$.  Note that the two steps within each cycle are not equally spaced but the pattern is periodic with period $T$. }
	\label{fig:example}
\end{figure}

Based on these same electrical components, a remarkable experiment by Cassidy {\it et al.}\cite{Cassidy} recently demonstrated the operation of a so-called Josephson junction laser.   The device is a resonator cavity (a half-wave coplanar waveguide) with a Josephson junction coupled to one end of the cavity near an antinode of the cavity electric field and biased with a DC voltage (See Fig.~\ref{fig:example}a).   While this general experimental configuration has been used in numerous experiments previously
(see for example \cite{Hofheinz,Wilson,Chen,Chen2,Gramich,Armour}), Ref.~\cite{Cassidy} observes for the first time many features of lasing. It is quite surprising that this effect has previously been overlooked.  Such a device, as a very narrow band controllable in-situ, or even on-chip, microwave source  could have great practical value for microwave applications, and therefore fully understanding its operation is essential.  In the context of  quantum computing applications,\cite{Cassidy,Devoret1,Devoret2,LesHouches}, such a device could provide a uniquely practical way to generate in-situ microwaves for switching transmon qubits without having to send microwave power externally into a cryostat, which is problematic for thermal isolation. 

In Ref.~\cite{Cassidy}, a theory to describe the newly observed phenomena was proposed (see also Refs.~\cite{Armour,Meister} where similar theoretical work is developed).   This theory gives a set of many simultaneous differential equations describing the many excitation modes of the resonator cavity.  The numerical solution of this system of equations reproduces much of the experiment.   Unfortunately, due to the complexity of this system of equations,  it is quite hard to develop much intuition or, without extensive numerical simulation, predict any of its properties.  The purpose of the current paper is to reformulate the physics in a much more transparent way to advance our understanding as well as our ability to accurately numerically simulate this type of experiment. 

%

To describe the dynamics of multiple excitation modes of a transmission line cavity we write equations for damped and driven oscillators:
\begin{equation}
\ddot \phi_n = -\omega_n^2  \phi_n - 2 \gamma \dot \phi_n +  \alpha _n F(t)
\label{eq:start}
\end{equation}
where  $\omega_n$ is the frequency of the $n^{th}$ cavity mode, $\gamma$ is the damping, $F(t)$ is a forcing function, and $\alpha_n$ is the coupling of the $n^{th}$ mode to the force.   In the case of a Josephson junction coupled to the cavity, biased with a DC voltage $V$, the forcing function is (up to a constant) the current injected by the junction\cite{Barone,Likharev}
\begin{equation}
  F(t) = \lambda \sin\left( \mbox{$ V t + \sum_n  \alpha_n \phi_n$} \right)
  \label{eq:start2}
\end{equation}
where $\lambda=2 E_J/C$ with $E_J$ the Josephson energy and $C$ the total capacitance of the waveguide (capacitance per unit length times length).     

The equations of motion (Eqs.~\ref{eq:start} and \ref{eq:start2}) have been previously derived in Supplemental Materials of Ref. \cite{Cassidy} and \cite{Armour}.  See also the Supplemental Material of the current paper\cite{supplement} for a detailed rederivation.    Note that $\phi_n$ represents the time-integrated voltage of the $n^{th}$ mode, and the argument of the $\sin$ is the superconducting phase difference across the junction. 
For an ideal waveguide $\omega_n = n \omega_0$ with $\omega_0$ the fundamental frequency of the cavity, $\alpha_n=1$ meaning all modes feel the same force,  and $\gamma=0$. However, these assumptions are not crucial (See Supplemental Material\cite{supplement}).   
In Ref.~\cite{Cassidy} lasing was found in numerical simulations  when many modes of the junction were considered  ($\alpha_n$ nonzero and approximately unity for $n$ up to about 20), thus giving a system of many simultaneous differential equations.   We will simplify this complicated system to a single equation of motion. 

The key to our analysis is to work in a real-time picture rather than in terms of the individual modes of the cavity.   We find solutions in which the many modes of the
cavity coherently combine to form discrete pulses which reflect back and forth in the cavity, as in a mode-locked laser, and which have a natural description in the time
domain.

We first think about the response of each mode to a driving force at the position of the junction. We denote the retarded Green's function of the $n^{th}$  simple harmonic oscillator as $G_n(t)$, which in the absence of damping takes the simple form $G^0_n(t) = \Theta(t) \sin(\omega_n t)/\omega_n$ with $\Theta$ the step function.   (The superscript ${}^0$ indicates no loss.) Since all of the modes couple to the same source, we group them all together by defining
\begin{equation}
 \Psi = \mbox{$\sum_n \alpha_n \phi_n $}\,.
 \label{eq:Psidef}
\end{equation}
The voltage across the junction is simply $V_J=V + \dot \Psi$.    The retarded Green's function for $\Psi$ is then just
$
 K(t) = \sum_n \alpha_n^2 G_n(t)
$. 
The dynamics of the system may then be recast as a single equation
\begin{equation}
 \Psi(t) = \lambda \int_{-\infty}^{t} \! \! dt'  K(t-t') \sin[Vt' + \Psi(t')]
 \label{eq:dynamics1}
\end{equation}
This is a highly nonlinear equation.  As is often the case with such equations, many solutions may exist and solutions may depend on initial conditions as well.

To understand the form of the kernel $K(t)$, consider first the ideal case of no damping $\gamma=0$ with $\omega_n = n \omega_0$ and $\alpha_n=1$.    Then $K$ has a sawtooth form $K^0(t) = (1/2)[T/2 - (t \mod T)]$ where $T=2\pi/\omega_0$ is the ``round-trip" time of the cavity.   Once one adds damping, with $\gamma \ll \omega_0$, to a very good approximation the response is the decaying sawtooth (See also Supplemental Material\cite{supplement})
\begin{equation}
\label{eq:decay}
K(t) = e^{-\gamma t} K^0(t)_.
\end{equation}
It is important to note that even if one adds randomness to the frequencies $\omega_n$ and to the couplings $\alpha_n$ the general sawtooth form persists (See also Supplemental Material\cite{supplement}). The sudden step reflects the fixed time delay associated with the round-trip time of the cavity.  


Note that in the case where $\Psi \ll 1$, which results from either small $\lambda T^2$ or large $\gamma T$, we can treat Eq.~\ref{eq:dynamics1} perturbatively.  At zeroth order, we drop $\Psi$ on the right hand side and have a simple integral on the right.   For example, in the case of large $\gamma T$, one obtains $\Psi^{(0)}(t) = (\lambda T/4 \gamma) \sin Vt$ with the superscript here meaning at zeroth order.   We can then plug this $\Psi^{(0)}$ into the right hand side of Eq.~\ref{eq:dynamics1} and again perform the integral, to obtain an improved approximation $\Psi^{(1)}$ at first order, and so forth.   It is easy to establish that this procedure only ever generates harmonics of the frequency $V$, i.e., the time period of oscillation is $2 \pi/V$.    At large $\gamma T$, this arises because the function $K(t)$ has decayed to almost zero before reaching $T$, so its sawtooth form has been lost and the time period $T$  forgotten.  

However, for small $\gamma T$ (believed to be appropriate for the experiment\cite{Cassidy}) with large $\lambda T^2$ there is a different type of solution where the oscillation period will instead be $T$.   Most of the remainder of this paper will explore this case.

Using the form of Eqs.~\ref{eq:decay}, we can transform Eq.~\ref{eq:dynamics1} to 
\begin{equation}
\Psi(t)  = \Psi(t-T) e^{-\gamma T} \!+ \lambda \int_{t-T}^{t} \!\!\!\!\! dt' K(t-t') \sin[Vt' + \Psi(t')]
\label{eq:dynamics2}
\end{equation}
The first term on the right represents the integration from $-\infty$ to $t-T$.   The interpretation is that a signal $\Psi(t-T)$ has gone down the waveguide and returned after time $T$ having decayed by $e^{-\gamma T}$.  The resulting signal $\Psi(t)$ is this decayed signal plus 
the result of driving by the Josephson junction during the period $t-T$ to $t$. 

Let us consider first the special case of $\gamma=0$ and a voltage $V$ that is commensurate with the period $T$.  I.e., we set $V=V_m = 2 \pi m/T$ for some integer $m$. Looking for a periodic solution we set $\Psi(t-T)=\Psi(t)$.    Eq.~\ref{eq:dynamics2} can then be satisfied by having the $\sin$ be a constant since the integral of $K^0$ over a period vanishes.  Thus we set
\begin{equation}
\label{eq:form}
 \Psi(t) \!\!\! \mod 2 \pi = -V_m t  + \beta
\end{equation}
for some constant $\beta$.  The right hand side is a linearly decreasing function of time, but can be made periodic by inserting $m$ phase slips of $2 \pi$ during the cycle (i.e., making it a sawtooth).  The phase slips may be at any point in the cycle of time $T$, although they need to be the same  from one cycle to the next to ensure $T$-periodicity.   This analytic solution matches numerical solutions for small $\gamma$ and commensurate $V$ quite well as shown, for example, in Fig.~\ref{fig:example}b.   This form of solution remains valid for any form  of $K^0(t)$ so long as its integral over a period vanishes.  In particular this will be true for any parameters $\alpha_n$ we choose in Eq.~\ref{eq:Psidef} (See also Supplemental Material\cite{supplement}).    We emphasize that this is the first completely analytic understanding of this experimental system.

\begin{figure}
	\vspace*{-.7in}	
	\hspace*{-34pt} \includegraphics[width=2.9in,angle=270]{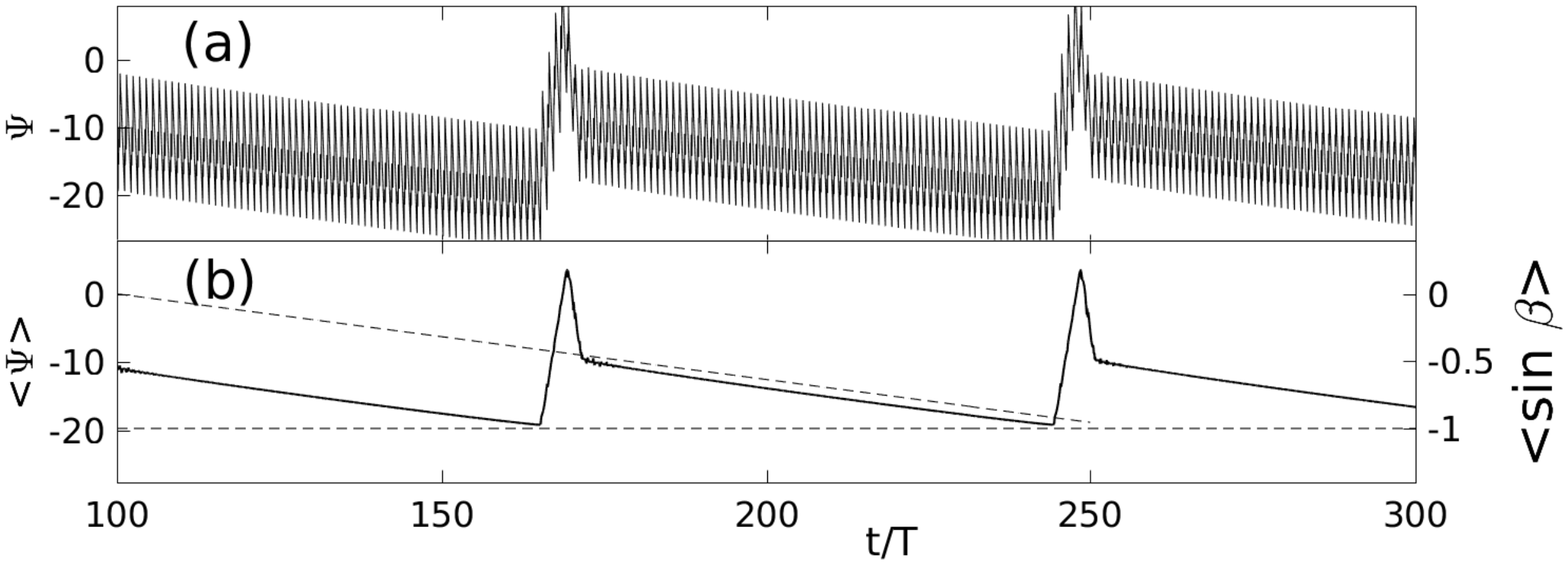}
	\vspace*{-.7in}
	\caption{Numerical form of $\Psi(t)$ obtained by integration of Eq.~\ref{eq:dynamics2}  for parameters $V = 4.02, \gamma=.05, \lambda=12$ in units where $T=2 \pi$.  (a) Oscillations of $\Psi$ over several hundred periods.  (b) Plot of $\langle \Psi \rangle$ which is $\Psi$ averaged over a period $T$ with vertical scale on the left, and also a plot of 
 $\langle \sin \beta \rangle$ which is $\sin[Vt + \Psi]$ averaged over a period $T$ with vertical scale on the right.  
  The two vertical scales are in ratio of $\lambda T^2/24$ as predicted by Eq.~\ref{eq:prop1}. 
    The two curves ($\langle \sin \beta \rangle$ and $\langle \Psi \rangle$) overlay so precisely that they are not both visible on this figure.  
   The diagonal dashed line is the predicted slope $-\delta V t$ as discussed in the text. Note that when $\sin \beta$ reaches $-1$ (the horizontal dashed line) the form of solution changes for a short period of time.}
	\label{fig:example2}
\end{figure}

\begin{figure}[h]
	\includegraphics[width=2in,angle=270]{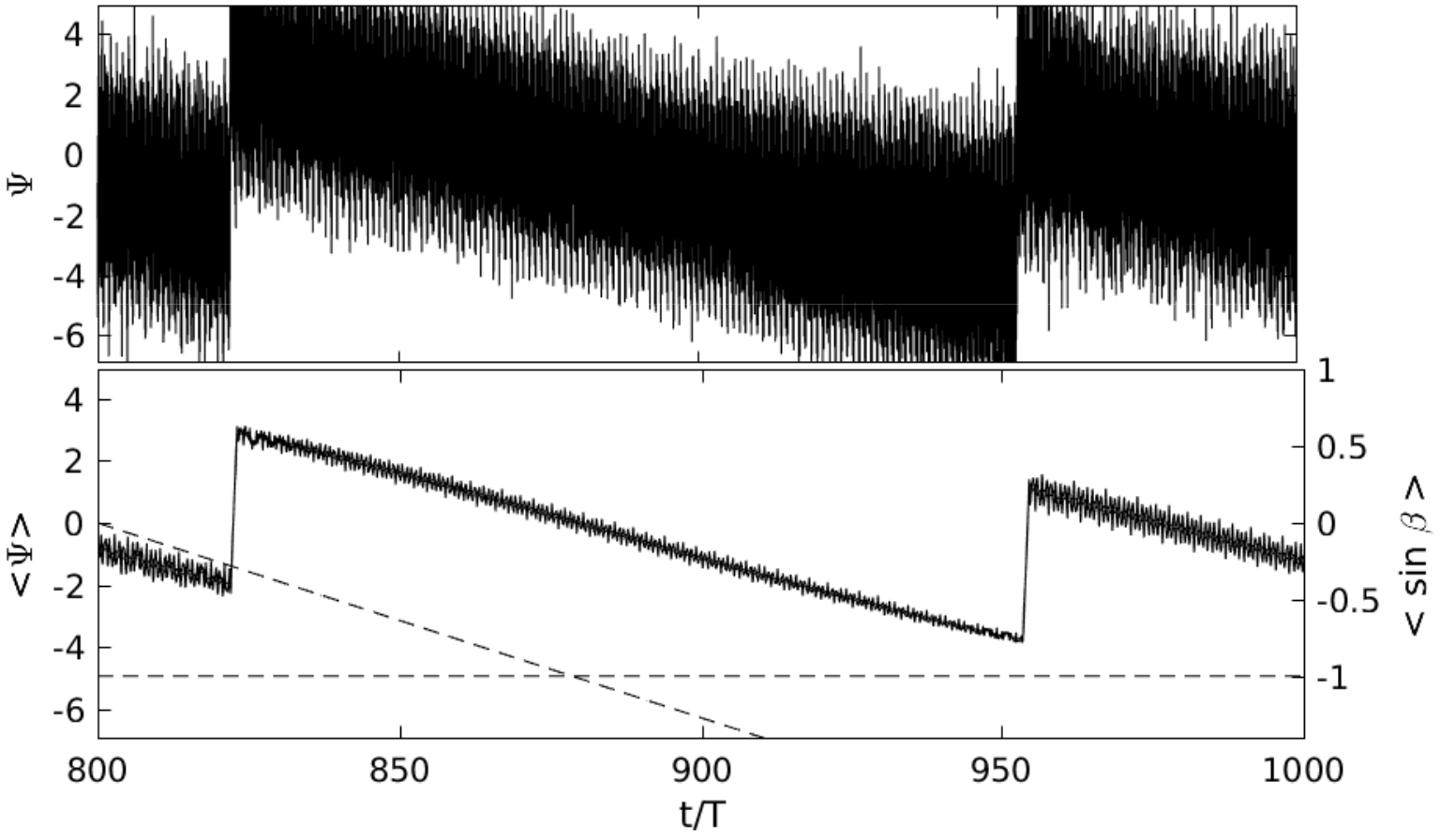} 
	\caption{As in Fig.~\ref{fig:example2} but for parameters closer to those of the experiment: $\lambda=3$, $\gamma=.0005$ and voltage $V=12.01$.  Again the plots of $\langle \Psi \rangle$ and $\langle \sin \beta \rangle$ overlap so one cannot distinguish the two curves.  More examples and details are given in the Supplemental Material\cite{supplement}. }
	\label{Fig:Psi3}
\end{figure}

The power absorbed by the waveguide is $P= (V_J - V) I =\dot \Psi E_J \sin(Vt + \Psi)$.  Here $V_J - V = \dot \Psi$ is the voltage from the injection point of the waveguide (the connection of the Josephson junction to the waveguide) to the  ground.   Thus the energy absorbed by the waveguide per cycle is 
$E_{\rm absorbed} = \int_{t-T}^t  dt' E_J \dot \Psi(t') \sin(Vt'  + \Psi(t'))  = -V T E_J \langle {\sin \beta} \rangle
$
where we have integrated by parts (and disregarded boundary terms assuming a periodic or almost periodic solution), and the brackets indicate the average of the $\sin$ over the cycle. 

If there is a damping $\gamma > 0$ then there will be a loss per cycle of $E_{\rm stored} (1-e^{- 2\gamma T})$ where $E_{\rm stored}$ is the stored energy in the waveguide.   For a steady state,  we  thus require that  $E_{\rm stored} =-V E_J \langle \sin \beta \rangle /(2\gamma)$, for small $\gamma T$.    From this we can estimate the maximum waveguide voltage (maximum $\dot \Psi$).  

Since $\Psi$ is basically a sum of sawtooth waves, we consider a sawtooth with a single $2 \pi$ step which has Fourier  modes  $\sin(2 \pi n t/T)$ of amplitude $2/n$.   To determine the energy stored in each of these modes, refer back to Eq.~\ref{eq:start} and note that the energy stored in a single oscillator is\footnote{Because the spatial form of the voltage is ${\cal V}_n(x) = V_n \cos(2 \pi n x/\ell)$, the total energy stored is half of the usual $CV^2/2$. See Supplemental Material\cite{supplement}}   
$
 E_{\rm stored} = \frac{C}{4} V_{\rm max}^2 = \frac{C}{4} (\dot \phi_n)^2 = \frac{C}{4} \left(\frac{2 \pi n}{T}\frac{2}{n}\right)^2 = \frac{4 \pi^2 C}{T^2}.
$
To represent the finite slope of the steps, we cut off the higher Fourier modes by using instead Fourier modes of amplitude $(2/n) f_n$ where $f_n$ is a some cutoff function which is unity at small $n$ and decays for large $n$.  The total energy stored is then $E_{\rm stored} = (4 \pi^2 C/T^2) 
\sum_n f_n^2$ whereas the maximum value of the voltage  will be  $V_{\rm max}=\dot \Psi_{\rm max} = (4 \pi/T) \sum_n f_n$. Given that there are $m = VT/(2 \pi)$ phase slips per cycle we add together all the energy in all of these pulses to get $E_{\rm stored} =  \xi C V V_{\rm max}/2$ where $\xi=\sum f_n^2/\sum_n f_n$.  For $f_n$ chosen as a sharp cutoff, $\xi=1$ whereas for an exponential cutoff  $f_n = e^{-a n}$ instead we obtain $\xi=1/2$.   Setting the energy stored to  $-V E_J \langle \sin \beta \rangle/(2\gamma)$  with $\lambda=2 E_J/C$ we obtain the result
$
|\dot \Psi_{\rm max}| = |V_{\rm max}| = \frac{\lambda}{2 \xi \gamma} |\langle \sin \beta \rangle|
$
with $\xi$ the unknown constant of order unity.  Numerically we find that this formula holds with $1/2 \lesssim \xi < 1$ whenever the solution has near to $T$-periodicity. 

With this maximum value of $\dot \Psi$ the sharp steps of $\Psi$ occur over a time-scale $\delta t \approx 2 \pi (2\xi\gamma)/(\lambda |\langle \sin \beta \rangle|)$.   During this time the argument of the $\sin$ wraps by $2 \pi$ and $\sin[Vt + \Psi]$ goes smoothly from  $\langle \sin \beta \rangle$ to $-1$ to $+1$ and then back to $\langle \sin \beta \rangle$.   The total area under this spike should then  be roughly $-2 \langle \sin \beta \rangle \delta t \approx 4 \pi \gamma/\lambda$.  (If $\langle \sin \beta \rangle=-1$ then it has to go all the way from $-1$ to $1$ giving a height of $2$ whereas if $\langle \sin \beta \rangle=0$ then the spike goes symmetrically up to $1$ and down to $-1$ having net zero area.) Thus we can approximate
\begin{equation}
\label{eq:steps2}
 \sin[Vt + \Psi(t)] = \langle \sin \beta \rangle + \mbox{$\sum_j$}  (4 \pi \gamma/\lambda) \delta(t-t_j)
\end{equation}
where $t_j$ are the particular times when the sawtooth steps occur.  This can then be plugged into Eq.~\ref{eq:dynamics1} and integrated.   The second term is responsible for producing the sawtooth steps in $\Psi$ since it is being integrated with the sawtooth function $K$.   Note that the coefficient of the delta function in Eq. \ref{eq:steps2} is exactly right to produce steps of size 2$\pi$ in $\Psi$.  On the other hand, when we integrate Eq.~\ref{eq:steps2} in Eq.~\ref{eq:dynamics1}, using $\int^t_{-\infty} dt' K(t-t') = T^2/24$, the first term gives the relationship
\begin{equation}
\label{eq:prop1}
 \langle \Psi \rangle = \lambda T^2 \langle \sin \beta \rangle /24
\end{equation}
where again the brackets mean an average over a full cycle.  This relationship is very accurately confirmed numerically, not only for the case of commensurate voltage that we have
focussed on so far, but also more generally, as shown in Figs.~\ref{fig:example2}b and \ref{Fig:Psi3}b.

We can now consider the case where $V$ deviates from being commensurate with the period $T$.  We write $V= V_m + \delta V$ where  $T \delta V  \ll 2\pi$.   We propose a solution of the form of Eq.~\ref{eq:form}  with $V_m$ still commensurate and now $\beta = -\delta V \,  t$ is a slow function of time.  We again assume there are $m$ phase slips of $2\pi$ per period and that they will occur at the same points in each cycle.  However now $\Psi$ is no longer periodic in $T$ but rather shifts by $-\delta V  T$ each cycle as seen in Figs.~\ref{fig:example2} and \ref{Fig:Psi3}.   The frequency of oscillation, however, remains  $\omega_0=2\pi/T$ very accurately.

Note that since $|\sin \beta|$ cannot exceed $1$, the form of solution must change after a time period $\sim \lambda T^2/(48 \delta V)$ as shown in the figure, before it locks back into quasi-periodic behavior. 

A few comments about the numerical solution of Eq.~\ref{eq:dynamics2}:  For small $\gamma T$, the $T$-periodic solution seems to be fairly stable.   As $\gamma$ increases, even for $\delta V=0$ (commensurate driving), $\beta$ does change slowly as a function of time, with a $\dot \beta$  that appears to be proportional to $\gamma^2 /(V  \lambda^2)$.   Also, for small $\gamma$, as $|\delta V|$ gets larger (i.e, for $V$ not very close to an integer multiple of $2\pi/T$), the assumption of small $T \delta V$ breaks down and the near-periodicity of the solution from one cycle to the next becomes imperfect.    We also note that for $\delta V$ small and negative a solution with $\Psi$ having the periodicity of $2 \pi/V$ rather than $T$ is fairly stable for $\gamma$ not too small.  We find numerically that for $|\delta V|  \gtrsim \gamma T$ the numerical solution reliably switches back to $T$-periodicity.    

Data is shown for parameters close to those of the experiment in Fig.~\ref{Fig:Psi3} with results analogous to those of Fig.~\ref{fig:example2}.   For these parameters, there is increased chaotic behavior although much similar physics is seen. See Supplemental Material\cite{supplement} for further details. 

In the actual experiment\cite{Cassidy}, the cavity $Q$ factor ($\sim$ 1000)  is 
mostly 
limited by the shunt conductance $G$ of the Josephson junction rather than the loss of the waveguide.  This conductance can be added into Eq.~\ref{eq:dynamics1} by adding a shunt\cite{Barone} current $G (V + \dot \Psi)$ to the Josephson current $E_J \sin[Vt + \Psi]$.   For small enough shunt conductance this does not significantly change the resulting dynamics.    Additional realistic effects can also be included in a similar way, such as the junction capacitance\cite{Barone} or changes to the Josephson current when the voltage is on the order of the gap.  Thus our approach has many parameters that might be tuned in an experiment, which are encoded in the precise form of the driving force $F$ as well as in the form of the response kernel $K$ as discussed above. 


One of the more remarkable observations of the experiment\cite{Cassidy}  is the phenomenon of injection locking behavior\cite{Adler,Petta}  reminiscent of laser physics.  While some of this physics is reproduced by our theory (See Supplemental material\cite{supplement} for details) there are  some features that we are not yet able to reproduce in detail. This remains a topic of current research. 
 
The general behavior of Eq.~\ref{eq:dynamics1} is extremely robust --- relatively independent of the chosen $\omega_n$, or $\alpha_n$ and even relatively independent of the form of the driving function, in that the  $\sin$ can be replaced with a wide range of periodic functions.  We give a more detailed study of this robustness in Supplemental Material\cite{supplement}.

Our description of the Josephson junction laser has been entirely classical, since the Josephson phase is treated as a classical dynamical variable. Quantum behaviour is encoded in the uncertainty relation, $\Delta N \Delta\Psi\geq 1/2$, with $\Delta N$ the number of Cooper pairs and $\Psi$ the phase across the junction. Setting $\Delta N = C \Delta V_J/(2e)$, with $C$ the total capacitance of the stripline, and working in terms of the relevant dimensionless voltage $\tilde{V}_J \equiv 2eV_J/(\hbar\omega_0)$, this becomes $\Delta {\tilde V}_J \Delta \Psi \geq 4 (e^2/h) Z_0$ where $Z_0 \equiv \sqrt{L/C}$ is the impedance of the stripline. For the parameters of the experiments\cite{Cassidy}, this is $\Delta {\tilde V}_J \Delta \Psi \geq 0.01$, showing that  quantum effects are expected to be small in the regimes of laser operation, with ${\tilde V}_J$ and $\Psi$ both of order unity or more.   While one may wonder to what extent it is appropriate to call a classical system a laser, we note that it has long been understood that the essential physics of a laser is recovered in classical physics\cite{classicallaser} and the operation of a free-electron laser is well-described by classical physics\cite{fel,felbook}. The stimulated emission into the cavity is fully accounted for by a classical description of the (nonlinear) drive.

Many of the qualitative features of the Josephson junction laser also appear in other nonlinear dynamical systems, perhaps the most familiar of which are musical instruments. A wide variety of sustained-tone musical instruments can be viewed as consisting of a linear resonator (e.g. violin string or organ pipe) which is subjected to a nonlinear drive (the violin bow, or organ reed). The nonlinearity of the drive establishes mode-locked oscillations of the linear resonator, leading to output at a fundamental and its pure harmonics even in situations in which the 
linear resonator itself is anharmonic\cite{fletcher,Jenkins}. For the same reason, the oscillations of the Josephson junction laser are relatively insensitive to anharmonicities in the waveguide. Another closely related system is provided by the Gunn oscillator, where the negative differential conductance provides a nonlinear drive of a microwave cavity, leading to stable oscillations with  similar mode-locked characteristics\cite{Tsai}.     There is however, a  feature that makes the equations of motion special for superconducting devices  ---the superconducting phase, the time integral of the voltage, being the physical quantity.  In particular this means that the argument of the nonlinear driving term includes $Vt$ and the resulting measured voltage is $\dot \Psi$.   Perhaps a better analogy is provided by a mode-locked superfluorescent optical laser\cite{Harvey}.   In Ref.~\onlinecite{Harvey} a laser coupled to a cavity dumper gives mode-locking associated with a round-trip time analogous to the current work.  There is, however, a difference that in Ref.~\onlinecite{Harvey}, the gain medium is extended rather than localized.

In summary we have presented an analytic framework for analysis of the Josephson Junction laser device.   The presence of a non-linear element coupled to a cavity gives mode-locked emission at the round-trip time of the cavity at least in some regimes.   Our time-domain framework greatly simplifies both numerical and analytic work and should aid further development of this field. 

{\bf Acknowledgements:} SHS is grateful to Lucas Casparis and the QDev journal club for introducing him in this problem.   SHS has been supported by the Niels Bohr International Academy, the Simons Foundation, and EPSRC Grants EP/I031014/1 and EP/N01930X/1. NRC is supported by EPSRC Grants EP/P034616/1 and EP/K030094/1.

\bibliography{jjbib2}

\begin{thebibliography}{24}%
\makeatletter
\providecommand \@ifxundefined [1]{%
 \@ifx{#1\undefined}
}%
\providecommand \@ifnum [1]{%
 \ifnum #1\expandafter \@firstoftwo
 \else \expandafter \@secondoftwo
 \fi
}%
\providecommand \@ifx [1]{%
 \ifx #1\expandafter \@firstoftwo
 \else \expandafter \@secondoftwo
 \fi
}%
\providecommand \natexlab [1]{#1}%
\providecommand \enquote  [1]{``#1''}%
\providecommand \bibnamefont  [1]{#1}%
\providecommand \bibfnamefont [1]{#1}%
\providecommand \citenamefont [1]{#1}%
\providecommand \href@noop [0]{\@secondoftwo}%
\providecommand \href [0]{\begingroup \@sanitize@url \@href}%
\providecommand \@href[1]{\@@startlink{#1}\@@href}%
\providecommand \@@href[1]{\endgroup#1\@@endlink}%
\providecommand \@sanitize@url [0]{\catcode `\\12\catcode `\$12\catcode
  `\&12\catcode `\#12\catcode `\^12\catcode `\_12\catcode `\%12\relax}%
\providecommand \@@startlink[1]{}%
\providecommand \@@endlink[0]{}%
\providecommand \url  [0]{\begingroup\@sanitize@url \@url }%
\providecommand \@url [1]{\endgroup\@href {#1}{\urlprefix }}%
\providecommand \urlprefix  [0]{URL }%
\providecommand \Eprint [0]{\href }%
\providecommand \doibase [0]{http://dx.doi.org/}%
\providecommand \selectlanguage [0]{\@gobble}%
\providecommand \bibinfo  [0]{\@secondoftwo}%
\providecommand \bibfield  [0]{\@secondoftwo}%
\providecommand \translation [1]{[#1]}%
\providecommand \BibitemOpen [0]{}%
\providecommand \bibitemStop [0]{}%
\providecommand \bibitemNoStop [0]{.\EOS\space}%
\providecommand \EOS [0]{\spacefactor3000\relax}%
\providecommand \BibitemShut  [1]{\csname bibitem#1\endcsname}%
\let\auto@bib@innerbib\@empty
\bibitem [{\citenamefont {Barone}\ and\ \citenamefont
  {Paterno}(1982)}]{Barone}%
  \BibitemOpen
  \bibfield  {author} {\bibinfo {author} {\bibfnamefont {A.}~\bibnamefont
  {Barone}}\ and\ \bibinfo {author} {\bibfnamefont {G.}~\bibnamefont
  {Paterno}},\ }\href {\doibase 10.1002/352760278X.fmatter} {\emph {\bibinfo
  {title} {Physics and Applications of the Josephson Effect}}}\ (\bibinfo
  {publisher} {Wiley-VCH Verlag GmbH \& Co. KGaA},\ \bibinfo {year}
  {1982})\BibitemShut {NoStop}%
\bibitem [{\citenamefont {Likharev}(1984)}]{Likharev}%
  \BibitemOpen
  \bibfield  {author} {\bibinfo {author} {\bibfnamefont {K.~K.}\ \bibnamefont
  {Likharev}},\ }\href@noop {} {\emph {\bibinfo {title} {The Dynamics of
  Josephson Junctions and Circuits}}}\ (\bibinfo  {publisher} {Gordon and
  Breach, New York},\ \bibinfo {year} {1984})\BibitemShut {NoStop}%
\bibitem [{\citenamefont {Devoret}\ \emph {et~al.}(2004)\citenamefont
  {Devoret}, \citenamefont {Wallraff},\ and\ \citenamefont
  {Martinis}}]{Devoret1}%
  \BibitemOpen
  \bibfield  {author} {\bibinfo {author} {\bibfnamefont {M.~H.}\ \bibnamefont
  {Devoret}}, \bibinfo {author} {\bibfnamefont {A.}~\bibnamefont {Wallraff}}, \
  and\ \bibinfo {author} {\bibfnamefont {J.~M.}\ \bibnamefont {Martinis}},\
  }\href@noop {} {\  (\bibinfo {year} {2004})},\ \Eprint
  {http://arxiv.org/abs/arXiv:cond-mat/0411174} {arXiv:cond-mat/0411174}
  \BibitemShut {NoStop}%
\bibitem [{\citenamefont {Devoret}\ and\ \citenamefont
  {Schoelkopf}(2013)}]{Devoret2}%
  \BibitemOpen
  \bibfield  {author} {\bibinfo {author} {\bibfnamefont {M.~H.}\ \bibnamefont
  {Devoret}}\ and\ \bibinfo {author} {\bibfnamefont {R.~J.}\ \bibnamefont
  {Schoelkopf}},\ }\href {\doibase 10.1126/science.1231930} {\bibfield
  {journal} {\bibinfo  {journal} {Science}\ }\textbf {\bibinfo {volume}
  {339}},\ \bibinfo {pages} {1169} (\bibinfo {year} {2013})}\BibitemShut
  {NoStop}%
\bibitem [{Les(2014)}]{LesHouches}%
  \BibitemOpen
  \href@noop {} {\emph {\bibinfo {title} {Quantum machines : measurement and
  control of engineered quantum systems}}},\ \bibinfo {series} {Ecole d'\'et\'e
  physique th\'eorique (Les Houches, Haute-Savoie, France)}, Vol.~\bibinfo
  {volume} {96}\ (\bibinfo  {publisher} {Oxford University Press},\ \bibinfo
  {year} {2014})\BibitemShut {NoStop}%
\bibitem [{\citenamefont {Cassidy}\ \emph {et~al.}(2017)\citenamefont
  {Cassidy}, \citenamefont {Bruno}, \citenamefont {Rubbert}, \citenamefont
  {Irfan}, \citenamefont {Kammhuber}, \citenamefont {Schouten}, \citenamefont
  {Akhmerov},\ and\ \citenamefont {Kouwenhoven}}]{Cassidy}%
  \BibitemOpen
  \bibfield  {author} {\bibinfo {author} {\bibfnamefont {M.~C.}\ \bibnamefont
  {Cassidy}}, \bibinfo {author} {\bibfnamefont {A.}~\bibnamefont {Bruno}},
  \bibinfo {author} {\bibfnamefont {S.}~\bibnamefont {Rubbert}}, \bibinfo
  {author} {\bibfnamefont {M.}~\bibnamefont {Irfan}}, \bibinfo {author}
  {\bibfnamefont {J.}~\bibnamefont {Kammhuber}}, \bibinfo {author}
  {\bibfnamefont {R.~N.}\ \bibnamefont {Schouten}}, \bibinfo {author}
  {\bibfnamefont {A.~R.}\ \bibnamefont {Akhmerov}}, \ and\ \bibinfo {author}
  {\bibfnamefont {L.~P.}\ \bibnamefont {Kouwenhoven}},\ }\href {\doibase
  10.1126/science.aah6640} {\bibfield  {journal} {\bibinfo  {journal}
  {Science}\ }\textbf {\bibinfo {volume} {355}},\ \bibinfo {pages} {939}
  (\bibinfo {year} {2017})},\ \Eprint
  {http://arxiv.org/abs/http://science.sciencemag.org/content/355/6328/939.full.pdf}
  {http://science.sciencemag.org/content/355/6328/939.full.pdf} \BibitemShut
  {NoStop}%
\bibitem [{\citenamefont {Hofheinz}\ \emph {et~al.}(2011)\citenamefont
  {Hofheinz}, \citenamefont {Portier}, \citenamefont {Baudouin}, \citenamefont
  {Joyez}, \citenamefont {Vion}, \citenamefont {Bertet}, \citenamefont
  {Roche},\ and\ \citenamefont {Esteve}}]{Hofheinz}%
  \BibitemOpen
  \bibfield  {author} {\bibinfo {author} {\bibfnamefont {M.}~\bibnamefont
  {Hofheinz}}, \bibinfo {author} {\bibfnamefont {F.}~\bibnamefont {Portier}},
  \bibinfo {author} {\bibfnamefont {Q.}~\bibnamefont {Baudouin}}, \bibinfo
  {author} {\bibfnamefont {P.}~\bibnamefont {Joyez}}, \bibinfo {author}
  {\bibfnamefont {D.}~\bibnamefont {Vion}}, \bibinfo {author} {\bibfnamefont
  {P.}~\bibnamefont {Bertet}}, \bibinfo {author} {\bibfnamefont
  {P.}~\bibnamefont {Roche}}, \ and\ \bibinfo {author} {\bibfnamefont
  {D.}~\bibnamefont {Esteve}},\ }\href {\doibase
  10.1103/PhysRevLett.106.217005} {\bibfield  {journal} {\bibinfo  {journal}
  {Phys. Rev. Lett.}\ }\textbf {\bibinfo {volume} {106}},\ \bibinfo {pages}
  {217005} (\bibinfo {year} {2011})}\BibitemShut {NoStop}%
\bibitem [{\citenamefont {Wilson}\ \emph {et~al.}(2011)\citenamefont {Wilson},
  \citenamefont {Johansson}, \citenamefont {Pourkabirian}, \citenamefont
  {Simoen}, \citenamefont {Johansson}, \citenamefont {Duty}, \citenamefont
  {Nori},\ and\ \citenamefont {Delsing}}]{Wilson}%
  \BibitemOpen
  \bibfield  {author} {\bibinfo {author} {\bibfnamefont {C.~M.}\ \bibnamefont
  {Wilson}}, \bibinfo {author} {\bibfnamefont {G.}~\bibnamefont {Johansson}},
  \bibinfo {author} {\bibfnamefont {A.}~\bibnamefont {Pourkabirian}}, \bibinfo
  {author} {\bibfnamefont {M.}~\bibnamefont {Simoen}}, \bibinfo {author}
  {\bibfnamefont {J.~R.}\ \bibnamefont {Johansson}}, \bibinfo {author}
  {\bibfnamefont {T.}~\bibnamefont {Duty}}, \bibinfo {author} {\bibfnamefont
  {F.}~\bibnamefont {Nori}}, \ and\ \bibinfo {author} {\bibfnamefont
  {P.}~\bibnamefont {Delsing}},\ }\href {\doibase 10.1038/nature10561}
  {\bibfield  {journal} {\bibinfo  {journal} {Nature}\ }\textbf {\bibinfo
  {volume} {479}},\ \bibinfo {pages} {376} (\bibinfo {year}
  {2011})}\BibitemShut {NoStop}%
\bibitem [{\citenamefont {Chen}\ \emph {et~al.}(2011)\citenamefont {Chen},
  \citenamefont {Sirois}, \citenamefont {Simmonds},\ and\ \citenamefont
  {Rimberg}}]{Chen}%
  \BibitemOpen
  \bibfield  {author} {\bibinfo {author} {\bibfnamefont {F.}~\bibnamefont
  {Chen}}, \bibinfo {author} {\bibfnamefont {A.~J.}\ \bibnamefont {Sirois}},
  \bibinfo {author} {\bibfnamefont {R.~W.}\ \bibnamefont {Simmonds}}, \ and\
  \bibinfo {author} {\bibfnamefont {A.~J.}\ \bibnamefont {Rimberg}},\ }\href
  {\doibase 10.1063/1.3573824} {\bibfield  {journal} {\bibinfo  {journal}
  {Applied Physics Letters}\ }\textbf {\bibinfo {volume} {98}},\ \bibinfo
  {pages} {132509} (\bibinfo {year} {2011})},\ \Eprint
  {http://arxiv.org/abs/http://dx.doi.org/10.1063/1.3573824}
  {http://dx.doi.org/10.1063/1.3573824} \BibitemShut {NoStop}%
\bibitem [{\citenamefont {Chen}\ \emph {et~al.}(2014)\citenamefont {Chen},
  \citenamefont {Li}, \citenamefont {Armour}, \citenamefont {Brahimi},
  \citenamefont {Stettenheim}, \citenamefont {Sirois}, \citenamefont
  {Simmonds}, \citenamefont {Blencowe},\ and\ \citenamefont {Rimberg}}]{Chen2}%
  \BibitemOpen
  \bibfield  {author} {\bibinfo {author} {\bibfnamefont {F.}~\bibnamefont
  {Chen}}, \bibinfo {author} {\bibfnamefont {J.}~\bibnamefont {Li}}, \bibinfo
  {author} {\bibfnamefont {A.~D.}\ \bibnamefont {Armour}}, \bibinfo {author}
  {\bibfnamefont {E.}~\bibnamefont {Brahimi}}, \bibinfo {author} {\bibfnamefont
  {J.}~\bibnamefont {Stettenheim}}, \bibinfo {author} {\bibfnamefont {A.~J.}\
  \bibnamefont {Sirois}}, \bibinfo {author} {\bibfnamefont {R.~W.}\
  \bibnamefont {Simmonds}}, \bibinfo {author} {\bibfnamefont {M.~P.}\
  \bibnamefont {Blencowe}}, \ and\ \bibinfo {author} {\bibfnamefont {A.~J.}\
  \bibnamefont {Rimberg}},\ }\href {\doibase 10.1103/PhysRevB.90.020506}
  {\bibfield  {journal} {\bibinfo  {journal} {Phys. Rev. B}\ }\textbf {\bibinfo
  {volume} {90}},\ \bibinfo {pages} {020506} (\bibinfo {year}
  {2014})}\BibitemShut {NoStop}%
\bibitem [{\citenamefont {Gramich}\ \emph {et~al.}(2013)\citenamefont
  {Gramich}, \citenamefont {Kubala}, \citenamefont {Rohrer},\ and\
  \citenamefont {Ankerhold}}]{Gramich}%
  \BibitemOpen
  \bibfield  {author} {\bibinfo {author} {\bibfnamefont {V.}~\bibnamefont
  {Gramich}}, \bibinfo {author} {\bibfnamefont {B.}~\bibnamefont {Kubala}},
  \bibinfo {author} {\bibfnamefont {S.}~\bibnamefont {Rohrer}}, \ and\ \bibinfo
  {author} {\bibfnamefont {J.}~\bibnamefont {Ankerhold}},\ }\href {\doibase
  10.1103/PhysRevLett.111.247002} {\bibfield  {journal} {\bibinfo  {journal}
  {Phys. Rev. Lett.}\ }\textbf {\bibinfo {volume} {111}},\ \bibinfo {pages}
  {247002} (\bibinfo {year} {2013})}\BibitemShut {NoStop}%
\bibitem [{\citenamefont {Armour}\ \emph {et~al.}(2013)\citenamefont {Armour},
  \citenamefont {Blencowe}, \citenamefont {Brahimi},\ and\ \citenamefont
  {Rimberg}}]{Armour}%
  \BibitemOpen
  \bibfield  {author} {\bibinfo {author} {\bibfnamefont {A.~D.}\ \bibnamefont
  {Armour}}, \bibinfo {author} {\bibfnamefont {M.~P.}\ \bibnamefont
  {Blencowe}}, \bibinfo {author} {\bibfnamefont {E.}~\bibnamefont {Brahimi}}, \
  and\ \bibinfo {author} {\bibfnamefont {A.~J.}\ \bibnamefont {Rimberg}},\
  }\href {\doibase 10.1103/PhysRevLett.111.247001} {\bibfield  {journal}
  {\bibinfo  {journal} {Phys. Rev. Lett.}\ }\textbf {\bibinfo {volume} {111}},\
  \bibinfo {pages} {247001} (\bibinfo {year} {2013})}\BibitemShut {NoStop}%
\bibitem [{\citenamefont {Meister}\ \emph {et~al.}(2015)\citenamefont
  {Meister}, \citenamefont {Mecklenburg}, \citenamefont {Gramich},
  \citenamefont {Stockburger}, \citenamefont {Ankerhold},\ and\ \citenamefont
  {Kubala}}]{Meister}%
  \BibitemOpen
  \bibfield  {author} {\bibinfo {author} {\bibfnamefont {S.}~\bibnamefont
  {Meister}}, \bibinfo {author} {\bibfnamefont {M.}~\bibnamefont
  {Mecklenburg}}, \bibinfo {author} {\bibfnamefont {V.}~\bibnamefont
  {Gramich}}, \bibinfo {author} {\bibfnamefont {J.~T.}\ \bibnamefont
  {Stockburger}}, \bibinfo {author} {\bibfnamefont {J.}~\bibnamefont
  {Ankerhold}}, \ and\ \bibinfo {author} {\bibfnamefont {B.}~\bibnamefont
  {Kubala}},\ }\href {\doibase 10.1103/PhysRevB.92.174532} {\bibfield
  {journal} {\bibinfo  {journal} {Phys. Rev. B}\ }\textbf {\bibinfo {volume}
  {92}},\ \bibinfo {pages} {174532} (\bibinfo {year} {2015})}\BibitemShut
  {NoStop}%
\bibitem [{\citenamefont {Simon}\ and\ \citenamefont
  {Cooper}(2017)}]{supplement}%
  \BibitemOpen
  \bibfield  {author} {\bibinfo {author} {\bibfnamefont {S.~H.}\ \bibnamefont
  {Simon}}\ and\ \bibinfo {author} {\bibfnamefont {N.~R.}\ \bibnamefont
  {Cooper}},\ }\href@noop {} {\bibfield  {journal} {\bibinfo  {journal}
  {Supplemental Material}\ } (\bibinfo {year} {2017})}\BibitemShut {NoStop}%
\bibitem [{Note1()}]{Note1}%
  \BibitemOpen
  \bibinfo {note} {Because the spatial form of the voltage is ${\protect \cal
  V}_n(x) = V_n \protect \qopname \relax o{cos}(2 \pi n x/\ell )$, the total
  energy stored is half of the usual $CV^2/2$. See Supplemental Material\cite
  {supplement}}\BibitemShut {NoStop}%
\bibitem [{\citenamefont {Adler}(1946)}]{Adler}%
  \BibitemOpen
  \bibfield  {author} {\bibinfo {author} {\bibfnamefont {R.}~\bibnamefont
  {Adler}},\ }\href {\doibase 10.1109/JRPROC.1946.229930} {\bibfield  {journal}
  {\bibinfo  {journal} {Proceedings of the IRE}\ }\textbf {\bibinfo {volume}
  {34}},\ \bibinfo {pages} {351} (\bibinfo {year} {1946})}\BibitemShut
  {NoStop}%
\bibitem [{\citenamefont {Liu}\ \emph {et~al.}(2015)\citenamefont {Liu},
  \citenamefont {Stehlik}, \citenamefont {Gullans}, \citenamefont {Taylor},\
  and\ \citenamefont {Petta}}]{Petta}%
  \BibitemOpen
  \bibfield  {author} {\bibinfo {author} {\bibfnamefont {Y.-Y.}\ \bibnamefont
  {Liu}}, \bibinfo {author} {\bibfnamefont {J.}~\bibnamefont {Stehlik}},
  \bibinfo {author} {\bibfnamefont {M.~J.}\ \bibnamefont {Gullans}}, \bibinfo
  {author} {\bibfnamefont {J.~M.}\ \bibnamefont {Taylor}}, \ and\ \bibinfo
  {author} {\bibfnamefont {J.~R.}\ \bibnamefont {Petta}},\ }\href {\doibase
  10.1103/PhysRevA.92.053802} {\bibfield  {journal} {\bibinfo  {journal} {Phys.
  Rev. A}\ }\textbf {\bibinfo {volume} {92}},\ \bibinfo {pages} {053802}
  (\bibinfo {year} {2015})}\BibitemShut {NoStop}%
\bibitem [{\citenamefont {Borenstein}\ and\ \citenamefont
  {Lamb}(1972)}]{classicallaser}%
  \BibitemOpen
  \bibfield  {author} {\bibinfo {author} {\bibfnamefont {M.}~\bibnamefont
  {Borenstein}}\ and\ \bibinfo {author} {\bibfnamefont {W.~E.}\ \bibnamefont
  {Lamb}},\ }\href {\doibase 10.1103/PhysRevA.5.1298} {\bibfield  {journal}
  {\bibinfo  {journal} {Phys. Rev. A}\ }\textbf {\bibinfo {volume} {5}},\
  \bibinfo {pages} {1298} (\bibinfo {year} {1972})}\BibitemShut {NoStop}%
\bibitem [{\citenamefont {Hopf}\ \emph {et~al.}(1976)\citenamefont {Hopf},
  \citenamefont {Meystre}, \citenamefont {Scully},\ and\ \citenamefont
  {Louisell}}]{fel}%
  \BibitemOpen
  \bibfield  {author} {\bibinfo {author} {\bibfnamefont {F.}~\bibnamefont
  {Hopf}}, \bibinfo {author} {\bibfnamefont {P.}~\bibnamefont {Meystre}},
  \bibinfo {author} {\bibfnamefont {M.}~\bibnamefont {Scully}}, \ and\ \bibinfo
  {author} {\bibfnamefont {W.}~\bibnamefont {Louisell}},\ }\href {\doibase
  http://dx.doi.org/10.1016/0030-4018(76)90283-2} {\bibfield  {journal}
  {\bibinfo  {journal} {Optics Communications}\ }\textbf {\bibinfo {volume}
  {18}},\ \bibinfo {pages} {413 } (\bibinfo {year} {1976})}\BibitemShut
  {NoStop}%
\bibitem [{\citenamefont {Saldin}\ \emph {et~al.}(2000)\citenamefont {Saldin},
  \citenamefont {Schneidmiller},\ and\ \citenamefont {Yurkov}}]{felbook}%
  \BibitemOpen
  \bibfield  {author} {\bibinfo {author} {\bibfnamefont {E.}~\bibnamefont
  {Saldin}}, \bibinfo {author} {\bibfnamefont {E.}~\bibnamefont
  {Schneidmiller}}, \ and\ \bibinfo {author} {\bibfnamefont {M.}~\bibnamefont
  {Yurkov}},\ }\href@noop {} {\emph {\bibinfo {title} {The Physics of Free
  Electron Lasers}}}\ (\bibinfo  {publisher} {Springer},\ \bibinfo {year}
  {2000})\BibitemShut {NoStop}%
\bibitem [{\citenamefont {Fletcher}(1999)}]{fletcher}%
  \BibitemOpen
  \bibfield  {author} {\bibinfo {author} {\bibfnamefont {N.~H.}\ \bibnamefont
  {Fletcher}},\ }\href {http://stacks.iop.org/0034-4885/62/i=5/a=202}
  {\bibfield  {journal} {\bibinfo  {journal} {Reports on Progress in Physics}\
  }\textbf {\bibinfo {volume} {62}},\ \bibinfo {pages} {723} (\bibinfo {year}
  {1999})}\BibitemShut {NoStop}%
\bibitem [{\citenamefont {Jenkins}(2013)}]{Jenkins}%
  \BibitemOpen
  \bibfield  {author} {\bibinfo {author} {\bibfnamefont {A.}~\bibnamefont
  {Jenkins}},\ }\href {\doibase https://doi.org/10.1016/j.physrep.2012.10.007}
  {\bibfield  {journal} {\bibinfo  {journal} {Physics Reports}\ }\textbf
  {\bibinfo {volume} {525}},\ \bibinfo {pages} {167 } (\bibinfo {year}
  {2013})},\ \bibinfo {note} {self-oscillation}\BibitemShut {NoStop}%
\bibitem [{\citenamefont {Tsai}\ \emph {et~al.}(1970)\citenamefont {Tsai},
  \citenamefont {Rosenbaum},\ and\ \citenamefont {MacKenzie}}]{Tsai}%
  \BibitemOpen
  \bibfield  {author} {\bibinfo {author} {\bibfnamefont {W.-C.}\ \bibnamefont
  {Tsai}}, \bibinfo {author} {\bibfnamefont {F.~J.}\ \bibnamefont {Rosenbaum}},
  \ and\ \bibinfo {author} {\bibfnamefont {L.~A.}\ \bibnamefont {MacKenzie}},\
  }\href {\doibase 10.1109/TMTT.1970.1127357} {\bibfield  {journal} {\bibinfo
  {journal} {IEEE Transactions on Microwave Theory and Techniques}\ }\textbf
  {\bibinfo {volume} {18}},\ \bibinfo {pages} {808} (\bibinfo {year}
  {1970})}\BibitemShut {NoStop}%
\bibitem [{\citenamefont {Harvey}\ \emph {et~al.}(1989)\citenamefont {Harvey},
  \citenamefont {Leonhardt}, \citenamefont {Drummond},\ and\ \citenamefont
  {Carter}}]{Harvey}%
  \BibitemOpen
  \bibfield  {author} {\bibinfo {author} {\bibfnamefont {J.~D.}\ \bibnamefont
  {Harvey}}, \bibinfo {author} {\bibfnamefont {R.}~\bibnamefont {Leonhardt}},
  \bibinfo {author} {\bibfnamefont {P.~D.}\ \bibnamefont {Drummond}}, \ and\
  \bibinfo {author} {\bibfnamefont {S.}~\bibnamefont {Carter}},\ }\href
  {\doibase 10.1103/PhysRevA.40.4789} {\bibfield  {journal} {\bibinfo
  {journal} {Phys. Rev. A}\ }\textbf {\bibinfo {volume} {40}},\ \bibinfo
  {pages} {4789} (\bibinfo {year} {1989})}\BibitemShut {NoStop}%
\end{thebibliography}%
\widetext

\clearpage


\pagebreak
\newpage
\begin{widetext}
\begin{center}
\textbf{\large Supplemental Materials: Theory of the Josephson Junction Laser}
\vspace*{10pt}
\end{center}
\end{widetext}

\setcounter{equation}{0}
\setcounter{figure}{0}
\setcounter{table}{0}
\setcounter{page}{1}
\makeatletter
\renewcommand{\theequation}{S\arabic{equation}}
\renewcommand{\thefigure}{S\arabic{figure}}
\renewcommand{\bibnumfmt}[1]{[S#1]}

\subsection{Supp. 1: Derivation of Eq.~\ref{eq:start}}

Although Eq.~\ref{eq:start} was given in Ref.~\cite{Cassidy} (see also Ref.~\cite{Armour}) and also for a single mode in \cite{Meister} we give a re-derivation of it here for completeness.  For simplicity we will perform the derivation in the absence of loss.  The inclusion of loss is quite straightforward. 

We begin with the telegrapher's equation for a waveguide.  Treating the system as a string of coupled inductors and capacitors we have the consitituitive equations
\begin{eqnarray}
  \partial_x  I &=& -{\cal C} \partial_t {\cal V}  \label{eq:tel1} \\
  \partial_x  {\cal V} &=& -{\cal L} \partial_t I  \label{eq:tel2}
\end{eqnarray}
where $\cal V$ is the electostatic potential, and $I$ is the current.   Here $\cal C$ is the capacitance per unit length and $\cal L$ is the inductance per unit length. One can view the system as being made of discrete 
elements separated by distance $a$, with ${\cal L}a$ and ${\cal C} a$ the 
inductance and capacitance of each individual element.

The boundary conditions for a half-wave cavity of length $\ell$ are $I(x=0)=I(x=\ell) = 0$ and correspondingly $\partial_x {\cal V}(x=0) = \partial_x {\cal V}(x=\ell) =0$.     Given these boundary conditions, we can write
\begin{eqnarray}
I &=& \sum_{n > 0} I_n(t) \sin(2 \pi n x/\ell) \\
{\cal V} &=&  \sum_{n \geq 0} {\cal V}_n(t) \cos(2 \pi n x/\ell) 
\end{eqnarray}

We now couple in the Josephson junction at position $x_i$. This injects current $I_{in}=E_j \sin(\varphi)$ at position $x_i$ where $\varphi$ is the phase across the junction (which we will determine later) and we have set $2e/\hbar=1$ as before.   Due to this current injection, we modify Eq. \ref{eq:tel1} to read
\begin{equation}
\label{eq:tel1a}
\partial_x  I = -{\cal C} \partial_t {\cal V}  + I_{in}(\varphi) \delta(x-x_i) 
\end{equation}
We can then decompose Eqns. \ref{eq:tel1a} and \ref{eq:tel2} into spatial Fourier modes giving (for $m > 0$)
\begin{eqnarray}
\label{eq:fir}
  (2 \pi m) I_m &=& -C \,  \partial_t {\cal V}_m + 2 \cos(2 \pi m x_i/\ell) I_{in}(\varphi) \\
  (2 \pi m) {\cal V}_m &=& L \,  \partial_t {I}_m  \label{eq:sec}
\end{eqnarray}
where $C={\cal C}\ell$ is the total capacitance and $L={\cal L} \ell$ the total inductance. 
Defining $$
\phi_m = \int_{-\infty}^{t} {\cal V}_m(t') dt'
$$
we have Eq.~\ref{eq:sec} written as
$$
  I_m =2 \pi m \phi_m/L 
$$
which we plug into  Eq. \ref{eq:fir} to obtain
\begin{equation}
\frac{(2 \pi m)^2}{L} \phi_m = -C   \partial_t^2 {\phi}_m + 2 \alpha_m I_{in}(\varphi) \label{eq:thir}
\end{equation}
where we have defined
\begin{equation}
\label{eq:alpham}
 \alpha_m = \cos(2 \pi m x_i/\ell).
\end{equation}
In the experiment, the injection point $x_i$ is very close to the end of the cavity so $\alpha_m \approx 1$ for $m$ not too large.  However, there is no reason not to consider the more general case. 

Looking also at the voltage at the injection point we have 
$$
 {\cal V}(x_{in})  =\sum_{m \geq 0} \alpha_m {\cal V}_m
$$
The waveguide is DC-biased using the techinque of Ref.~\cite{Chen}, which couples to the ${\cal V}_0$ mode which we then identify as the applied voltage $V$.  We thus have 

$$
 \varphi(t) = \int^t_{-\infty} dt' {\cal V}(x_{in},t') = V t + \sum_{m>0} \alpha_m \phi_m(t)
$$
Plugging this into Eq.~\ref{eq:thir} obtains our final result Eq.~\ref{eq:start}, except for the loss term which we have dropped only for convenience of notation here.

\subsection{Supp. 2: The Green's Function}

Given a damped harmonic oscillator with a $\delta$ function source
\begin{equation}
\ddot \phi= -\omega^2  \phi- 2 \gamma \dot \phi + \delta(t),
\end{equation}
it is easy to  show that the response is
$$
G(t) = \frac{e^{-\gamma t} \sin \tilde \omega t}{\tilde \omega}  \Theta(t) 
$$
where 
$$\tilde \omega = \sqrt{\omega^2 - \gamma^2}.$$
Now defining
$$
\Psi = \sum_n \alpha_n \phi_n 
$$
we get a response function for $\Psi$ given by 
\begin{equation}
K(t) = \sum_n  \alpha_n^2 \frac{e^{-\gamma t} \sin \tilde \omega_n t}{\tilde \omega_n}  \Theta(t).
\end{equation}

\begin{figure}[h]
\hspace*{-.5in}	\includegraphics[width=2in]{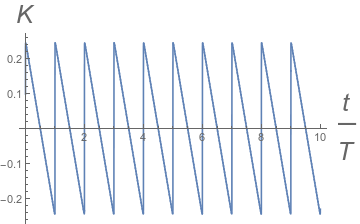} 
	\caption{This the ideal sawtooth $K^0$ with $\alpha_n=1$ and $\omega_n = n \omega_0$ and no loss, $\gamma=0$.}
	\label{fig:K0}
\end{figure}

For the simplest case discussed in the text ($\alpha_n=1, \gamma=0$), the form of $K$ is a simple sawtooth as shown in Fig.~\ref{fig:K0}.  

\begin{figure}[h]
	\hspace*{-.5in}	\includegraphics[width=2in]{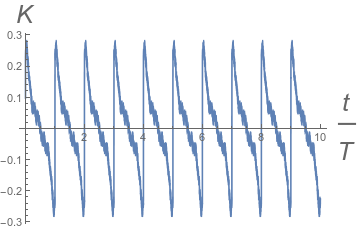} 
	\caption{This is $K(t)$ if $\alpha_n$ is chosen randomly between $.8$ and $1.2$ and $\gamma=0$.  Notice that the step features remain robust, and the periodicity is fully maintained.}
		\label{Fig:K1}
\end{figure}

\begin{figure}[h]
	\hspace*{-.5in}	\includegraphics[width=2in]{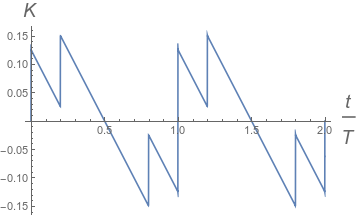} 
	\caption{This is $K(t)$ if $\alpha_n$ is given by Eq.~\ref{eq:alpham} with the injection point being at $x_i = .1 \ell$ and $\gamma=0$.  The multiple steps correspond to the time delays given by the round trip times to either end of the cavity.}
		\label{Fig:K3}
\end{figure}

In the case where couplings $\alpha_n$ have some randomness, it is important to point out that the periodicity remains strictly $T$, although the waveform changes.   Nonetheless, as shown in Fig.~\ref{Fig:K1} the sharp step rise remains quite robust.  
If $\alpha_n$ is smoothly cut off above some scale of $n$ (which appears to be what was used in the simulation of Ref.~\cite{Cassidy}), then the form of $K$ will be smooth, and the slope of the step will be limited by the cutoff.   A more physical situation is given by $\alpha_n$ given by Eq.~\ref{eq:alpham} corresponding to the Josephson junction not being positioned quite at the end of the cavity.  In this case there will be peaks in $K(t)$ corresponding to the reflection times from either end of the cavity.  An example of this is shown in Fig.~\ref{Fig:K3}.  The sawtooth response of $\Psi$ will reflect these multiple steps. 

In the case with $\gamma$ nonzero, as mentioned in the text a very good approximation is to take $K(t)=e^{-\gamma t} K^0(t)$.   The difference between this form and the actual calculated $K$ is due to the difference between $\omega_n = n \omega_0$ and $\tilde \omega_n$ which is tiny. Strictly speaking the difference is order $\gamma^2$ but plotting the two different functions for $\gamma$ even as large as unity, they are essentially indistinguishable. 

\begin{figure}[h]
	\hspace*{-.5in}	\includegraphics[width=2in]{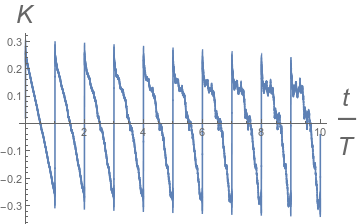} 
	\caption{This is $K(t)$ if $\omega_n$ is chosen randomly between $(n-.05)\omega_0$ and $(n+.05) \omega_0$ and $\gamma=0$.  While the step features remain robust, the periodicity is slowly destroyed.}
		\label{Fig:K2}
\end{figure}

Adding randomness to the frequencies $\omega_n$ does have a strong effect on $K$ as shown in Fig.~\ref{Fig:K2}, although the sharp step remains a robust feature. 

\subsection{Supp. 3: Numerical Results For Parameters Close to That of Experiment}

The parameters chosen for the figures in the main text are used mainly for clarity of presentation and to elucidate some of the physics that can occur. (Note that as mentioned in the main text, with nonlinear equations, multiple types of solutions may be possble.)   Much of the physics  we discuss in the main text is also observed in the more experimentally relevant parameter regime.  In the experiment of Cassidy {\it et al.} [1] the voltage range of lasing is between $V=6\omega_0-16 \omega_0$.  The $Q$-factor is stated as roughly $1000$, so $\gamma=0.0005$ since $Q=\omega_0/(2 \gamma)$.   The coupling constant is estimated as $\lambda \gtrsim 1$ in units where $T=2 \pi$. We take $\lambda=3$ as an example. 

\begin{figure}[h]
	\includegraphics[width=2.5in]{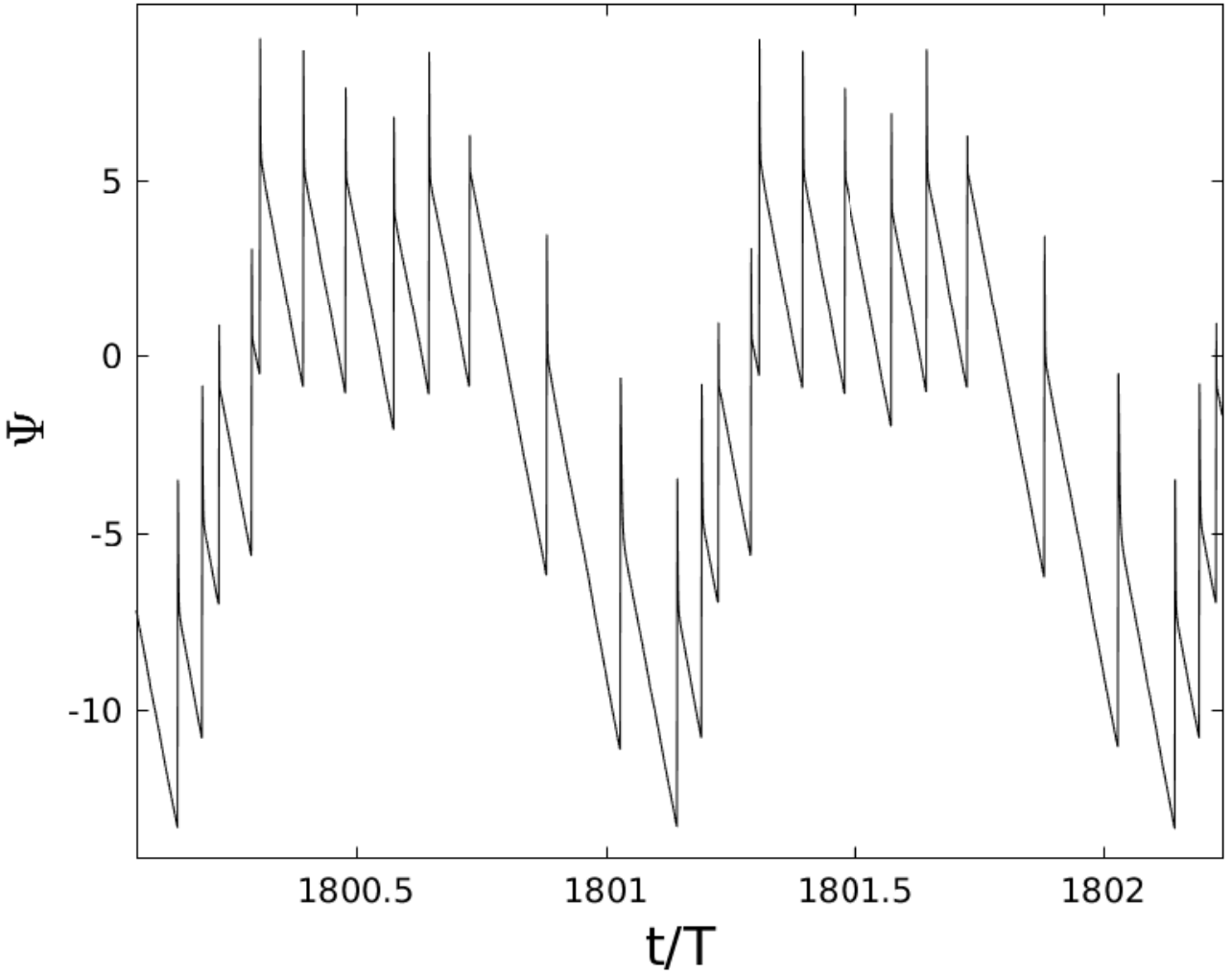} 
	\caption{Pattern of $\Psi$ for commensurate voltage with parameters close to that of the experiment.  With units where $T=2\pi$ we use $\lambda=3$ with $\gamma=.0005$ and commensurate driving $V=12.0$.  The sawtooth structure is evident although not perfect. }
	\label{Fig:Psi1}
\end{figure}

With units of $T=2\pi$ or $\omega_0=1$, we start by looking at a commensurate case of $V=12$ in Fig.~\ref{Fig:Psi1}.   Here again we see the characteristic sawtooth shape analogous to that of Fig.~\ref{fig:example}.  Note however, that the sawtooth here is not perfect.  

\begin{figure}[h]
	\includegraphics[width=2.5in]{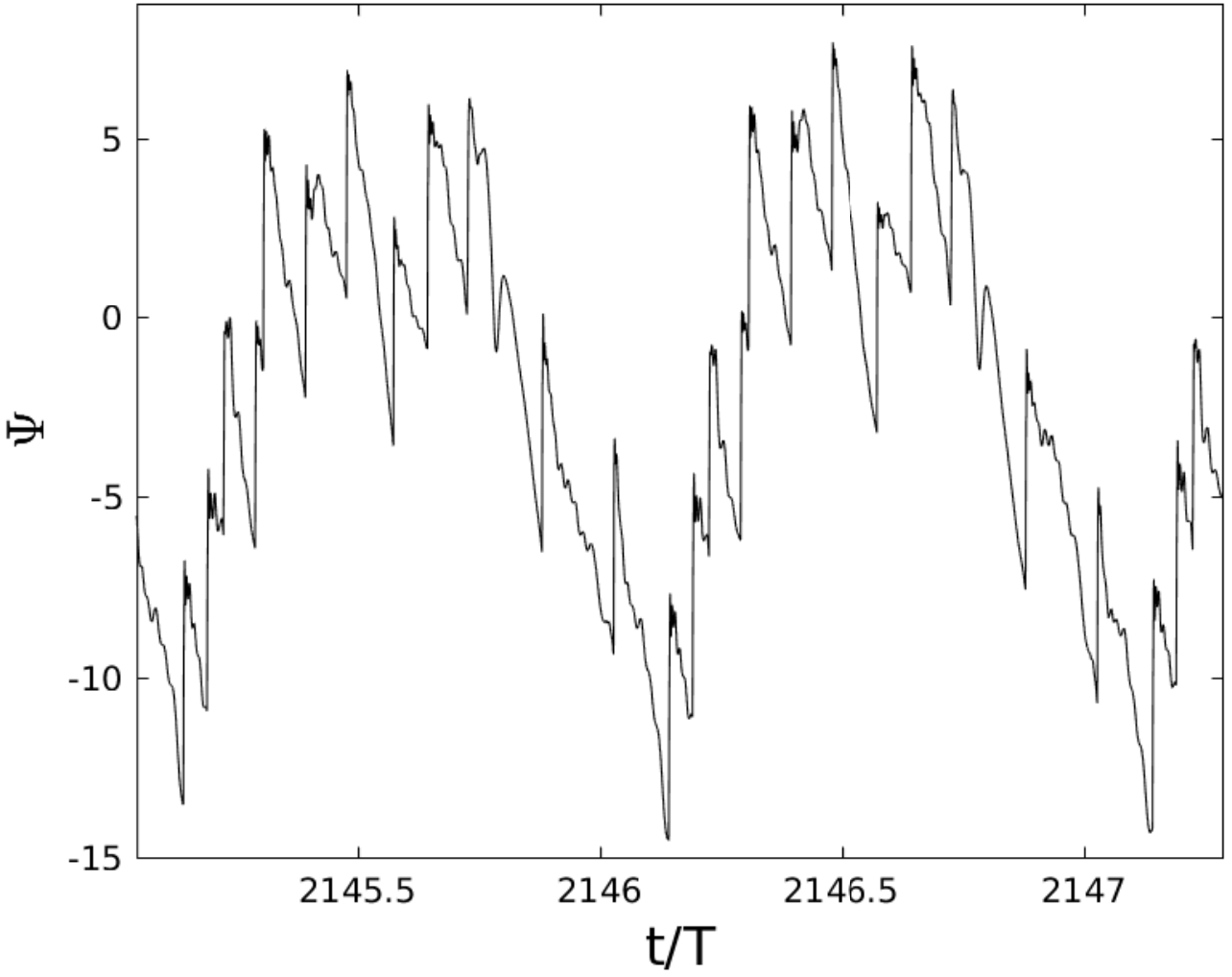} 
	\caption{Pattern of $\Psi$ with parameters close to that of the experiment.  With units where $T=2\pi$, we use $\lambda=3$ with $\gamma=.0005$ and commensurate driving $V=12.0$. }
	\label{Fig:Psi2}
\end{figure}

Other solutions to the nonlinear equation, which are even less sawtooth-like, may also appear as shown in Fig.~\ref{Fig:Psi2}.  Again there is still some sawtooth-like behavior, and clearly some periodicity, although obviously the signal is more complex. 

Moving slightly away from commensuration with $V=12.01$ we see in Fig.~\ref{Fig:Psi3} of the main text much of the same features as we have in Fig.~\ref{fig:example2} in the main text.  Note that here the slope is very slightly different from prediction, but otherwise seems to fit well and as with the main text $\langle \Psi\rangle$ and $\langle \sin \beta \rangle$ overlay exactly on top of each other.   Note that there are some imperfections in the periodicity of the oscillations in the top half of the figure.  


\begin{figure}[h]
	 \vspace*{10pt}
	\includegraphics[width=3in]{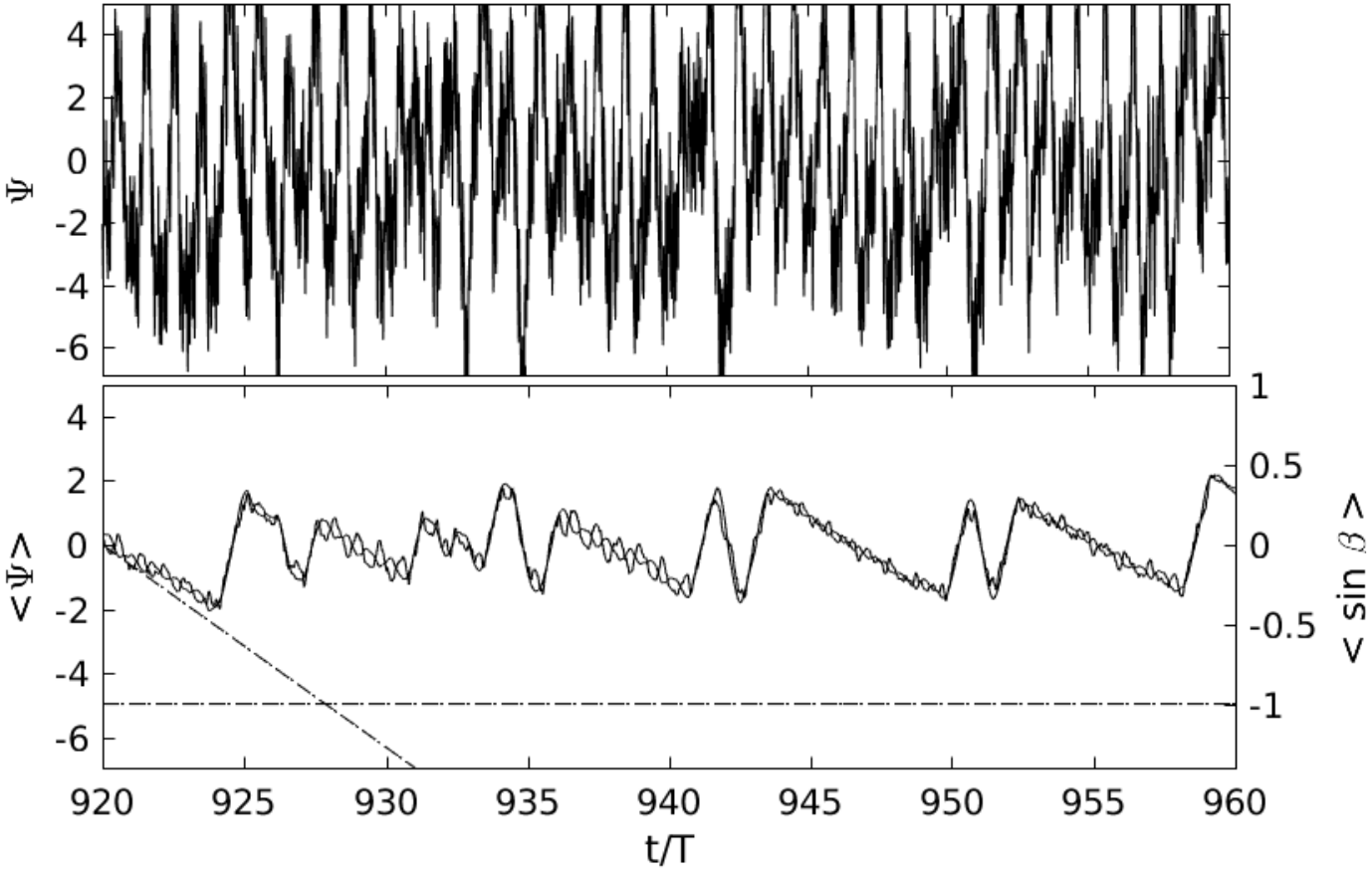} 
	\caption{Pattern of $\Psi$ for commensurate voltage with parameters close to that of the experiment.  With units where $T=2\pi$, we use $\lambda=3$ with $\gamma=.0005$ and voltage $V=12.1$. }
	\label{Fig:Psi4}
\end{figure}

 \begin{figure}[h]
	\includegraphics[width=3in]{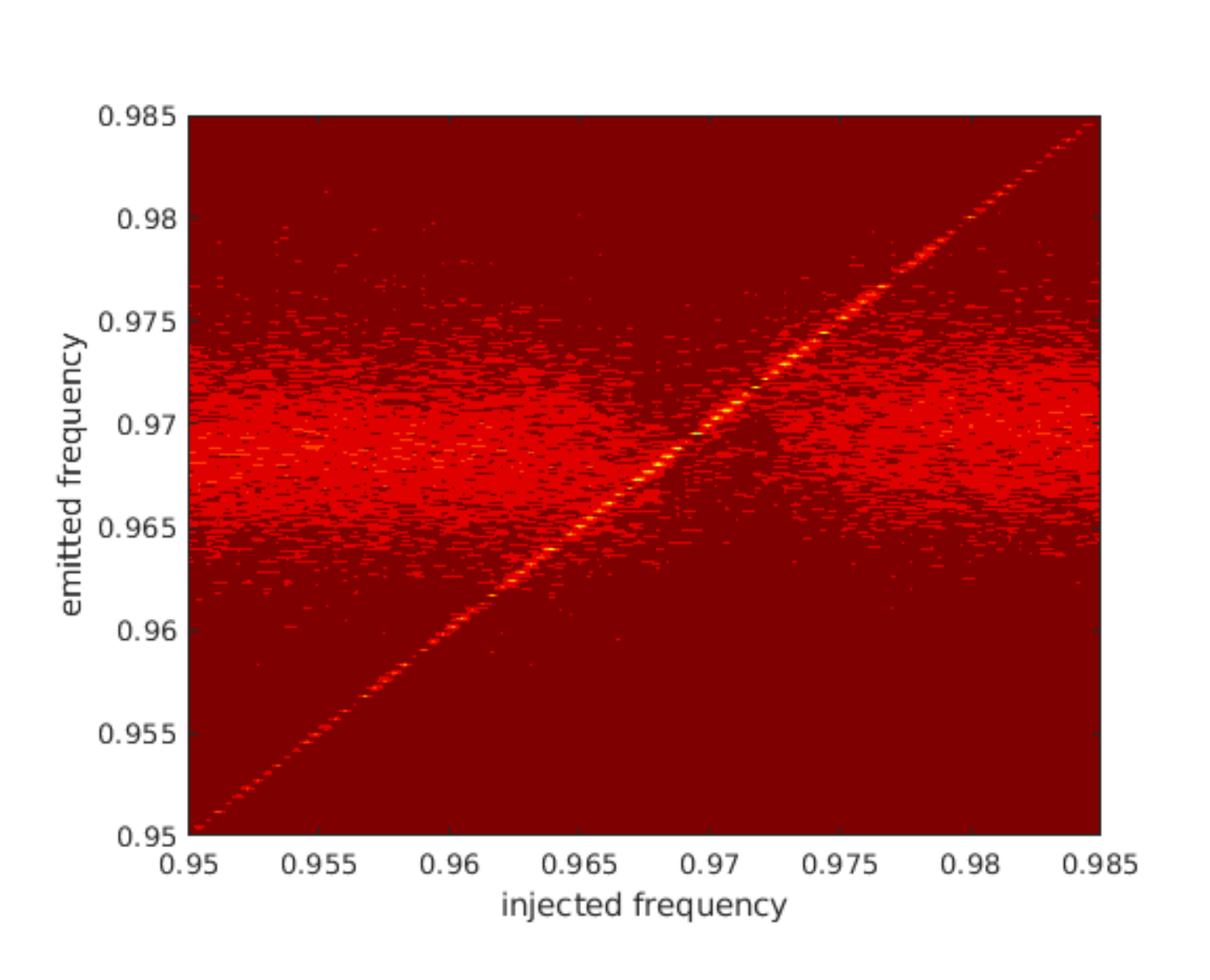} 
	\includegraphics[width=3in]{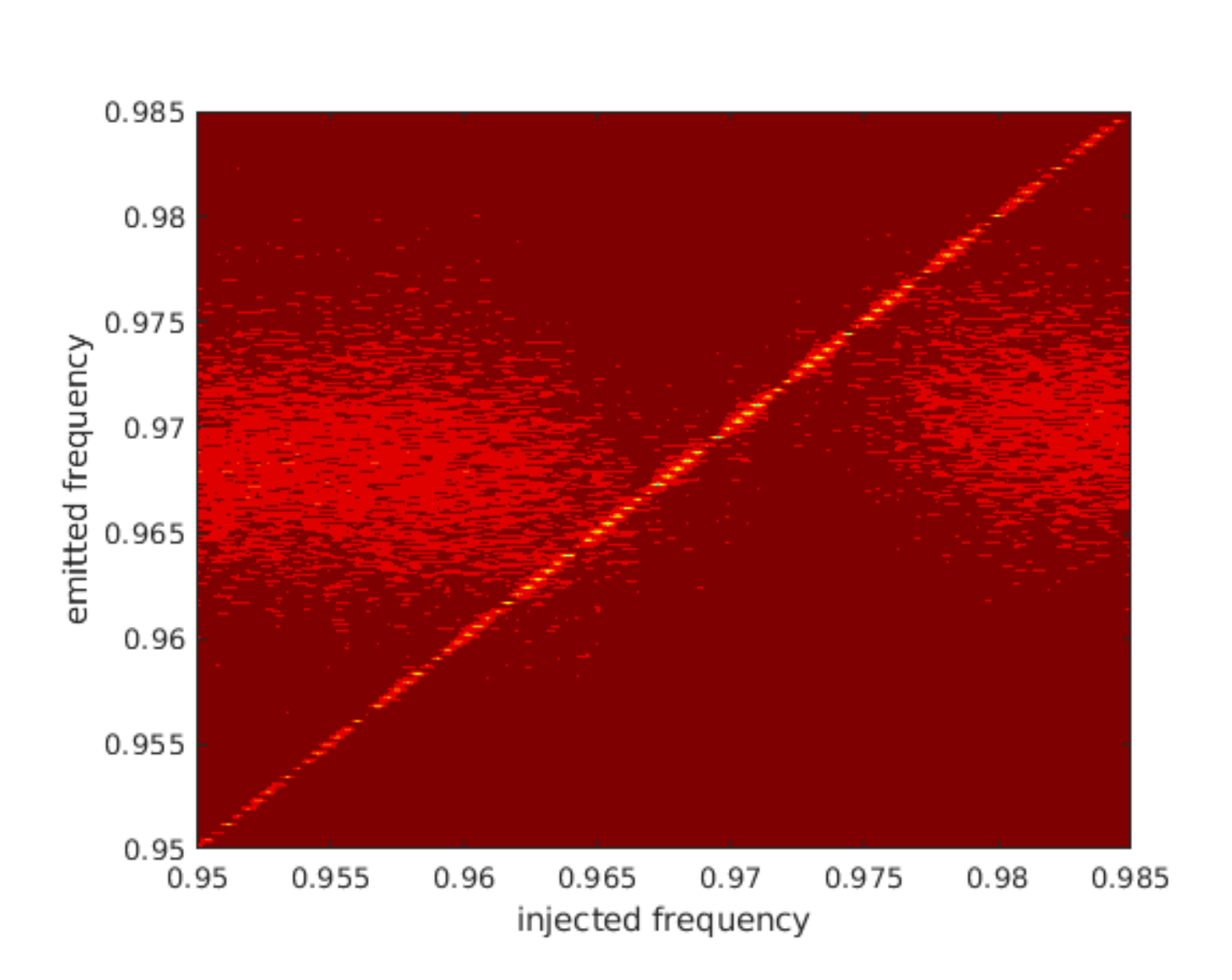} 
	\caption{Injection locking behavior.  At nonzero voltage, a weak signal is injected into the cavity and the emitted frequencies are observed.   In these figures, the natural emission frequency is approximately 0.9675, and has a spread in frequency of about 0.005.  When the injected signal is far enough from the natural frequency, it does not appear to influence the emitted signal.  On the other hand, when the injected signal is close enough the natural emission band, it locks the emission onto the injected frequency and the only frequency seen to emit is the same as that of the injected frequency --- this is seen as the large gap in the center of the plot with only the diagonal line passing through it.  The bottom figure has a 50\% stronger injected signal than the upper figure, and correspondingly, the range of frequencies for which the emission is locked to the injeced signal is increased.  In these figures $V=4.41, T=2 \pi, \lambda = 10, \gamma=0.05$.  The waveguide is given dispersion by using $\omega_n = n \omega_0 (.998)^{n-1}$. In the top figure the injected signal current is of amplitude $0.5$ and in the lower figure it is of amplitude $0.75$.  	  }
	\label{Fig:inject}
\end{figure}

Moving slightly further away from commensuration with $V=12.1$ we see in Fig.~\ref{Fig:Psi4}  much of the same features as we have in Fig.~\ref{fig:example2} in the main text and as in Fig.~\ref{Fig:Psi3} although the pattern is more complicated with a higher number of regions where the quasiperiodicity is lost.  Note that in the lower half of this figure one can discern the two separate curves for $\langle \Psi \rangle$ and $\langle \sin \beta \rangle$ although they overlap very closely.   Note that the average of $\langle \sin \beta \rangle$ does not strike $-1$ before the pattern makes a jump here.

\subsection{Supp 4: Injection Locking Behavior}

With the DC voltage applied, when an additional weak signal is injected into the cavity with a frequency within the range of the natural line width, the emitted signal sharpens and matches the injected frequency.  The well-known detailed theory of oscillator locking behavior by \cite{Adler} seems to match the experimental work of \cite{Cassidy}.   One might expect that our Equation of motion Eq.~\ref{eq:dynamics1} would display this locking behavior.   We have found that the current theory (Eq.~\ref{eq:dynamics1}) reproduces part of this phenomenology, but not all of it.  

We first spread the emission peak by giving some dispersion to the waveguide.  We then find that an injected signal can indeed sharpen the natural emission line when the injected frequency is within the natural emission band.  Further, in agreement with Refs.~\cite{Cassidy,Adler} we find that the frequency range over which injection locking occurs is increased when the injected power is increased as shown in Fig.~\ref{Fig:inject}.

In order to spread out the emission peak we add some dispersion to the cavity by using a simple model $\omega_n = n \omega_0 x^{n-1}$ with $x = .998$  (the random ``detuning" used by Ref.~\cite{Cassidy} has similar effect) .   The natural emitted frequency band is now at a frequency of about $.9675 \omega_0$ with a bandwidth of about $.0.005$ due to this dispersion.  Injecting a tone at frequency $\Omega$ close to $\omega_0$ generates a resonantly strong response of amplitude $A \sim \omega_0/(\omega_0 - \Omega)$.  This can be included in the dynamical equation Eq.~\ref{eq:dynamics1} by shifting $\Psi \rightarrow \Psi + A \cos(\Omega t)$.    By a redefinition of variables this is most easily included by inserting the term $A \cos(\Omega t)$ inside the $\sin$ of Eq.~\ref{eq:dynamics1}.   The equation of motion is the integrated numerically, with results shown in Fig.\ref{Fig:inject}.   Here the spectral weight of the broadened peak locks onto the injected frequency when the injected frequency is close enough to the natural emission frequency.  As in the theory of Adler\cite{Adler} and in the experiment, the stronger the injected signal, the wider a range of frequencies will lock the emitted signal as shown in Fig.  \ref{Fig:inject}.   Despite this similarity to the experiment\cite{Cassidy} and to the theory of Adler\cite{Adler}, there do seem to be  some differences.   For example in those works additional resonances are seen when the injected frequency is near to the emitted frequency, but not near enough to fully lock the emission.   We have not managed to replicate this phenomenon, and this remains a subject of current research.

\subsection{Supp 5: Varying Number of Modes}

\begin{figure}[h]
	\includegraphics[width=3.5in]{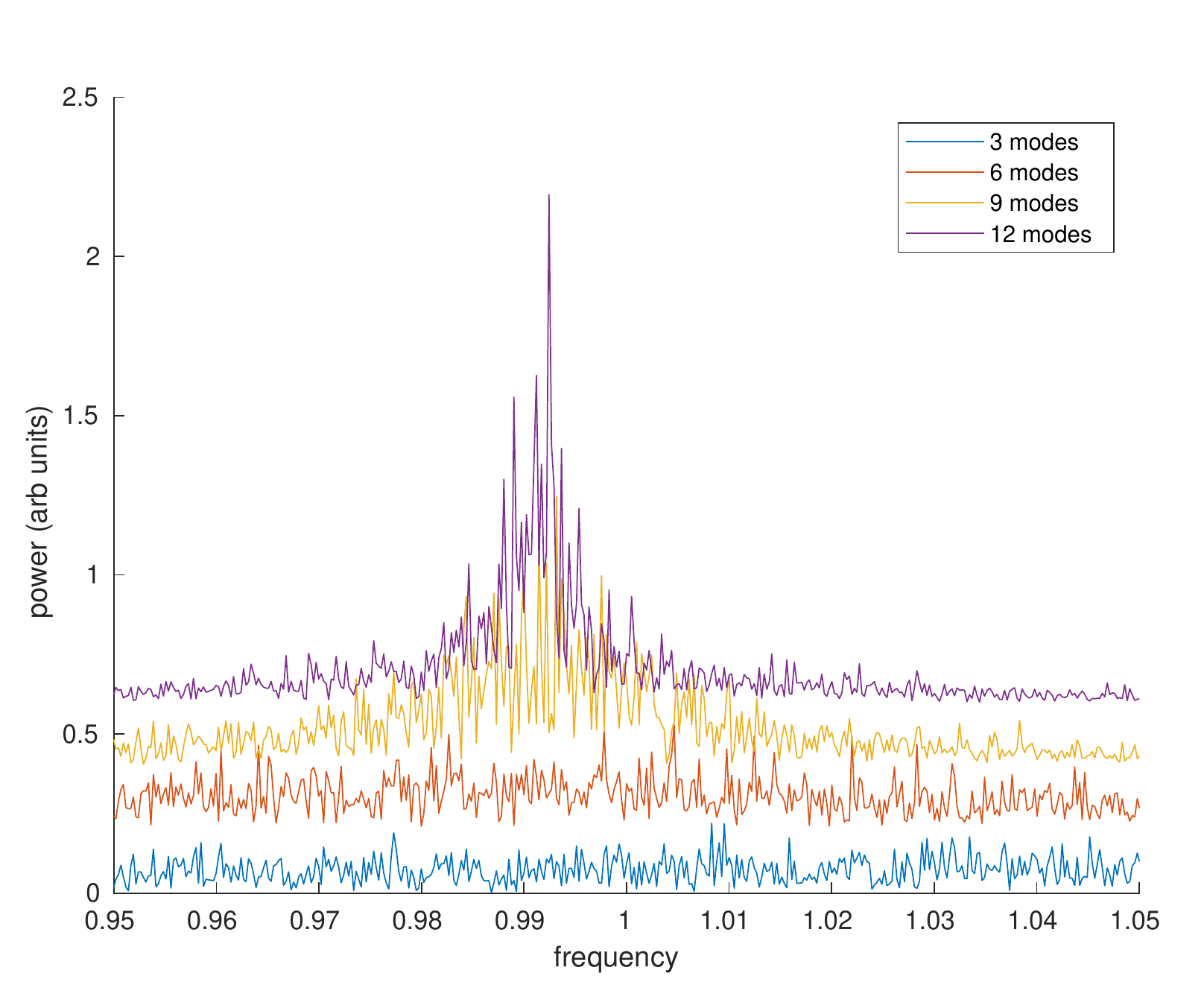} 
	\caption{Power spectrum of $\Psi$ with different number of modes of the cavity.    In order to have a sharp response at the frequency corresponding to the round trip time of the cavity, one needs roughly 10 Fourier modes.  Here we use $\omega_n = n \omega_0$ and $\alpha_n = 1$ for $n \leq$ number of modes.    With units where $T=2\pi$, we use $\lambda=10$ with $\gamma=0.05$ and voltage $V=2.41$.  The plots are offset for clarity. }
	\label{fig:modes}
\end{figure}

As in the theoretical work in the supplemental material of Ref.~\cite{Cassidy}, it is crucial to keep enough modes of the cavity in order to obtain a strong emission.   In the language of the current paper the interpretation of this is fairly simple --- it requires many Fourier modes to have a sharp sawtooth response function $K(t)$.      If too few modes are included, the response function does not have a sharp step and there is no strong emission at the fundamental frequency of the cavity.  In Fig.~\ref{fig:modes} we show that roughly 10 modes are required before a strong emission peak forms at the frequency corresponding to the round trip time of the cavity. 

\subsection{Supp 6: Varying The Current Injection Function}

\begin{figure}[h]
	\includegraphics[width=3.5in]{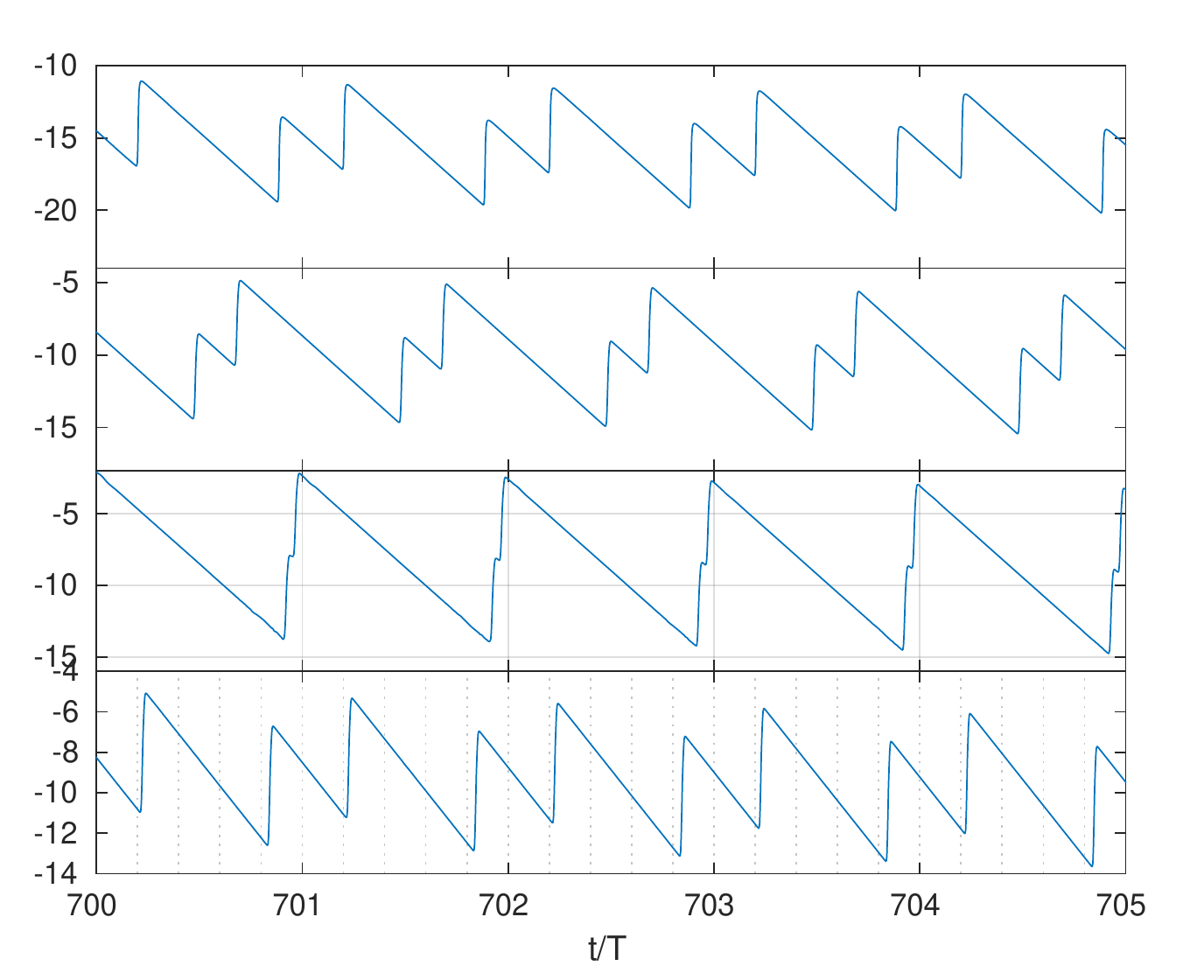} 
	\caption{Here we vary the current injection function $\sin(Vt + \Psi)$ to $[\sin(Vt + \Psi)]^n$.  From top to bottom, $n=1,3,5,7$.  We see that the physics is essentially unchanged.  In all cases, the signal is a sum of two sawtooth waves of amplitude $2 \pi$ with arbitrary temporal position in the cycle --- and the period is always $T$.   Here using units where $T=2\pi$, we use $\lambda=10$ with $\gamma=0.05$ and voltage $V=2.04$
	}
	\label{fig:sinpow}
\end{figure}

As mentioned in the text, the general response of the system is surprisingly robust to various changes in many details.   As an example in Fig.~\ref{fig:sinpow} we vary the current injection function.  Instead of using $\sin(V t + \Psi)$ we use $[\sin(V t + \Psi)]^n$.  As can be seen there, the result is essentially unchanged.  

\begin{figure}[h]
	\includegraphics[width=3.5in]{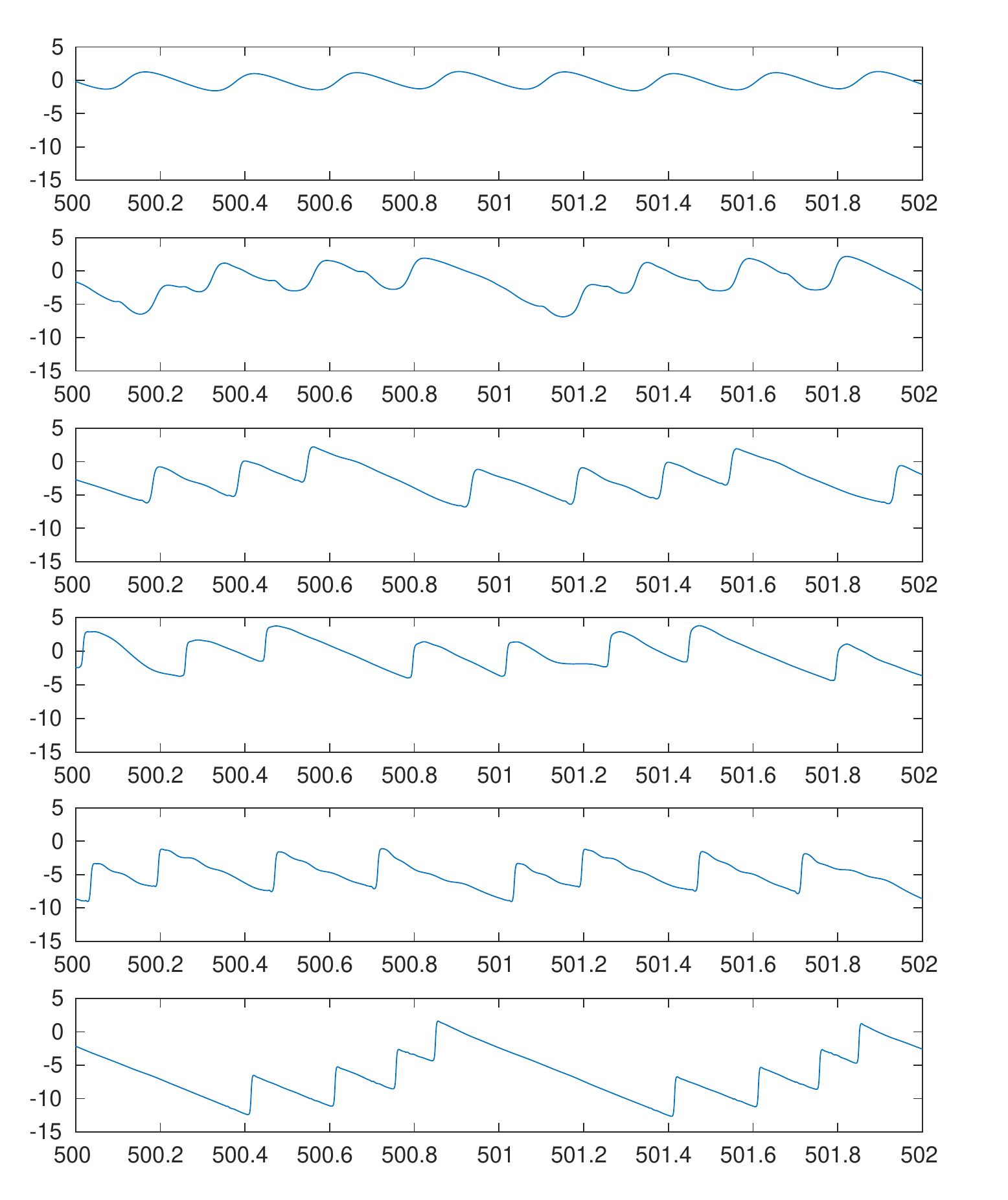} 
	\caption{$\Psi$ with different values of the coupling constant $\lambda$.    With units where $T=2\pi$, top to bottom we have $\lambda = .75, 1.5, 2.25, 3, 3.75, 4.25$  and voltage $V=4.04$ and $\gamma=0.01$.  }
	\label{fig:lamvary}
\end{figure}

\begin{figure}[h]
	\hspace*{10pt}\includegraphics[width=3.4in]{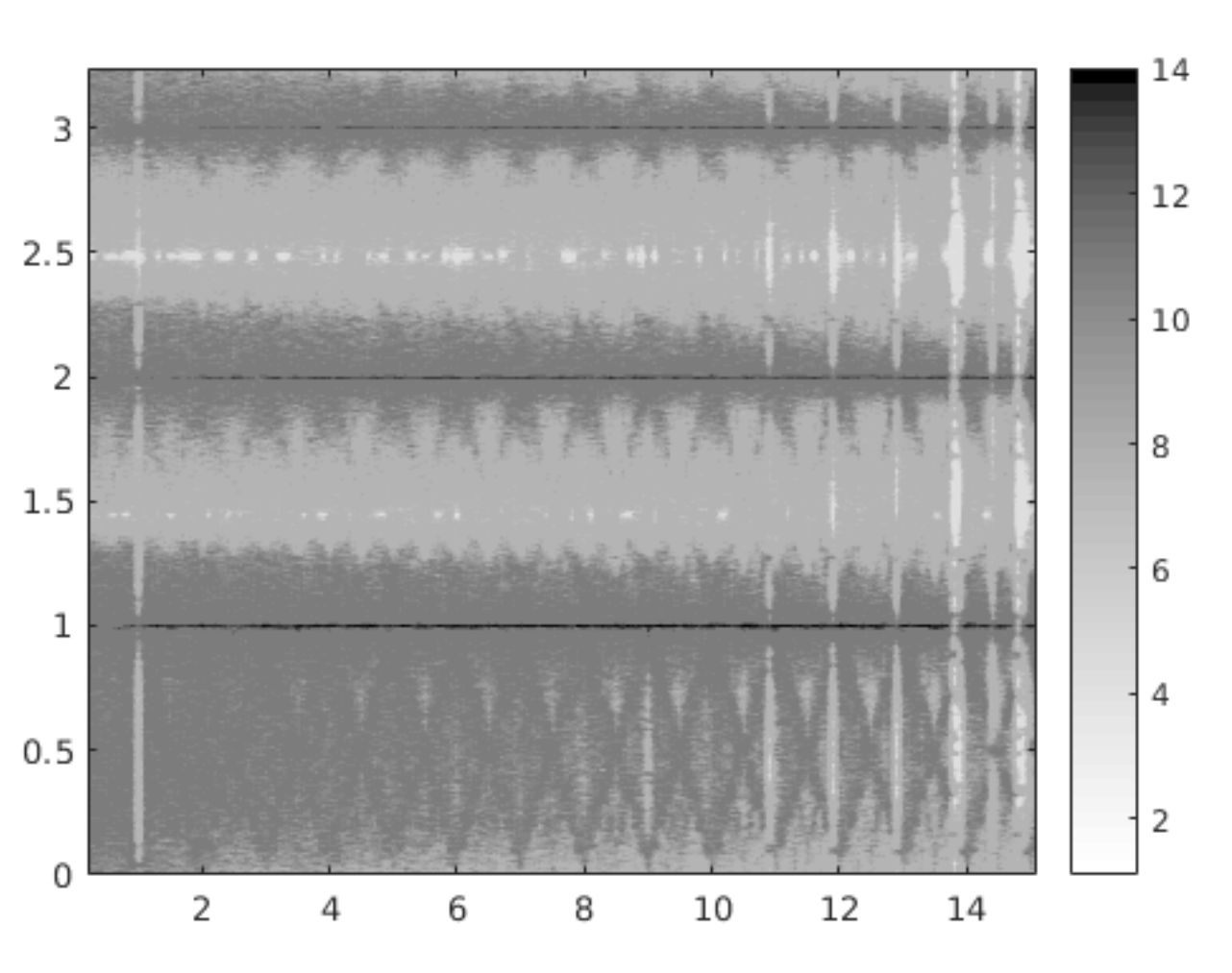} 
	
	\includegraphics[width=3.5in]{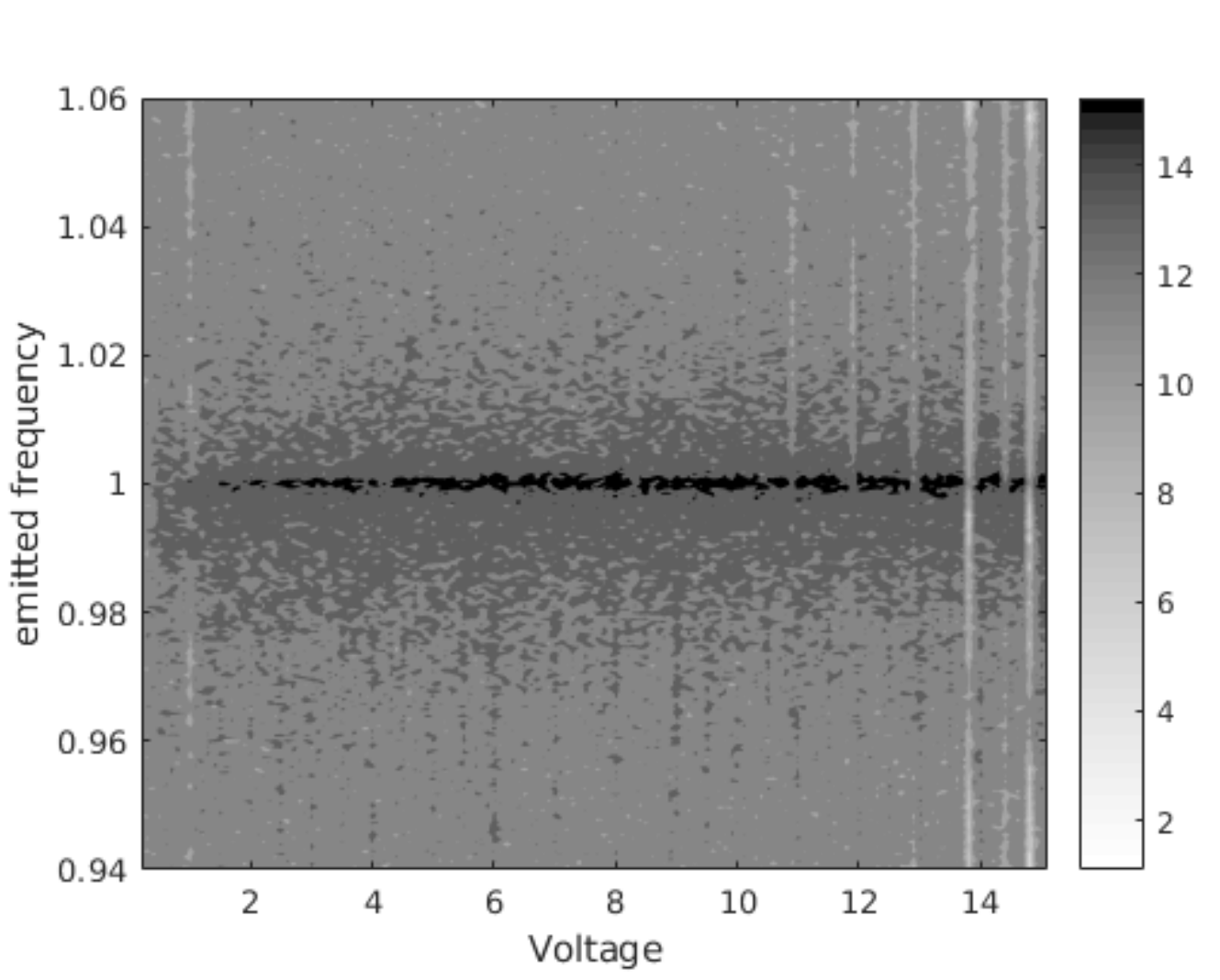} 
	\caption{Power spectrum of $\Psi$ for coupling constant $\lambda=2.46$ as in Ref.~\citep{Cassidy}.    Here $\omega_n = n \omega_0$ and $\gamma = 10^{-4}$.   The shaded graph intensity is log of the power in the Fourier spectrum of $\Psi$.  Lower plot is a zoom of the upper plot. }
	\label{fig:cassmatch}
\end{figure}

\subsection{Supp 7: Varying The Coupling $\lambda$}

We can consider varying the nonlinear coupling $\lambda$ as shown in Fig.~\ref{fig:lamvary}.    At weak coupling, as discussed in the text, one obtains oscillations at frequency $V$.  As $\lambda$ is increased, one sees harmonics of $V$. At large enough coupling (here somewhere around $\lambda$ approximately $4$) the solution strongly locks into the saw-tooth solution discussed in the text.

\subsection{Supp 8: Comparison to simulations of Cassidy et al.\cite{Cassidy}}

In Fig.~\ref{fig:cassmatch} we show a few plots using some of the same parameters as used for simulations in the supplemental material of Ref.~\cite{Cassidy}.   While the data is not identical to that of Ref.~\cite{Cassidy} (it need not be identical, given that this is a nonlinear equation that depends on initial conditions and may also be chaotic), it has many similarities --- particularly the strong peak at integer multiples of the fundamental frequency of the cavity.   Some additional features are visible in the upper plot that were not visible in Ref.~\cite{Cassidy} --- in particular one can see peaks in intensity at frequencies $m \omega_0 \pm V$ which show up as diamonds in the upper plot (mostly visible at frequencies below $\omega_0$ and at high voltage).

\clearpage


\end{document}